# Wavelet methods in statistics: Some recent developments and their applications[*]

## Anestis Antoniadis[†]

*Laboratoire Jean Kuntzmann*
*BP 53, 38041 Grenoble Cedex 9*
*France*
*e-mail:* `Anestis.Antoniadis@imag.fr`

**Abstract:** The development of wavelet theory has in recent years spawned applications in signal processing, in fast algorithms for integral transforms, and in image and function representation methods. This last application has stimulated interest in wavelet applications to statistics and to the analysis of experimental data, with many successes in the efficient analysis, processing, and compression of noisy signals and images.

This is a selective review article that attempts to synthesize some recent work on "nonlinear" wavelet methods in nonparametric curve estimation and their role on a variety of applications. After a short introduction to wavelet theory, we discuss in detail several wavelet shrinkage and wavelet thresholding estimators, scattered in the literature and developed, under more or less standard settings, for density estimation from i.i.d. observations or to denoise data modeled as observations of a signal with additive noise. Most of these methods are fitted into the general concept of regularization with appropriately chosen penalty functions. A narrow range of applications in major areas of statistics is also discussed such as partial linear regression models and functional index models. The usefulness of all these methods are illustrated by means of simulations and practical examples.

**AMS 2000 subject classifications:** Primary 60K35, 60K35; secondary 60K35.
**Keywords and phrases:** curve smoothing, density estimation, wavelet thresholding, penalized least-squares, robust regression, partial linear models, mixed effects models, inverse regression, time series prediction.

Received January 2007.

## Contents

---

[*]This paper was accepted by T. Tony Cai, Associate Editor for the IMS.
[†]Financial support form the IAP research network Nr. P5/24 and P6/03 of the Belgian government (Belgian Science Policy) is gratefully acknowledged. The author would also like to thank the Associate Editor and the two referees for their useful comments and suggestions.

## 1. Introduction

*Nonparametric regression*  has been a fundamental tool in data analysis over the past two decades and is still an expanding area of ongoing research. The goal is to recover an unknown function, say $g$, based on sampled data that are contaminated with noise. Denoising techiques provide a very effective and simple way of finding structure in data sets without the imposition of a parametric regression model (as in linear or polynomial regression for example). Only very general assumptions about $g$ are made such as that it belongs to a certain class of functions.

During the 1990s, the nonparametric regression literature was arguably dominated by (nonlinear) *wavelet shrinkage* and *wavelet thresholding* estimators. These estimators are a new subset of an old class of nonparametric regression estimators, namely orthogonal series methods. Moreover, these estimators are easily implemented through fast algorithms so they are very appealing in practical situations. Donoho and Johnstone (1994) and Donoho et al. (1995) have introduced nonlinear wavelet estimators in nonparametric regression through thresholding which typically amounts to term-by-term assessment of estimates of coefficients in the empirical wavelet expansion of the unknown function. If an estimate of a coefficient is sufficiently large in absolute value – that is, if it exceeds a predetermined threshold – then the corresponding term in the empirical wavelet expansion is retained (or shrunk toward to zero by an amount equal to the threshold); otherwise it is omitted.

Extensive reviews and descriptions of the various classical and Bayesian wavelet shrinkage and wavelet thresholding estimators are given in the books by Ogden (1997), Vidakovic (1999) and Percival and Walden (2000), in the papers appeared in the edited volume by Müller and Vidakovic (1999), and in the review papers by Antoniadis (1997), Vidakovic (1998b) and Abramovich et al. (2000). With the increased applicability of these estimators in nonparametric regression, several new wavelet based curve smoothing procedures have been proposed in the recent literature, and one of the purposes of this review is to present few of them under the general concept of penalized least squares regression. It should be noted that most of these methods are usually implemented under the assumptions of dyadic sample size, equally spaced and fixed sample points, and i.i.d. normal errors. When the data does not meet one or both of these requirements, various modifications have been proposed and we will also mention few of them in the forthcoming sections. To keep the length of this article reasonable we have not included in our discussion important developments in Bayesian wavelet denoising methods which deserve a survey by their own.

Finally we provide a brief overview of a spectrum of wavelet applications in some real data problems. The reader should be cautioned, however, that the wavelet is so a large research area that truly comprehensive surveys are almost impossible, and thus, our overview may be a little restricted. At first we will be concerned with a semiparametric partially linear regression model with unknown regression coefficients and we will present a wavelet thresholding based estimation procedure to estimate the components of the partial linear model. We will also discuss a wavelet based variant of a popular dimension reduction method for functional data analysis that is traditionally used in practice, namely MAVE.

The rest of the paper is organized as follows. Section 2 recalls some known results about wavelet series, function spaces and the discrete wavelet transform. Section 3 briefly discusses the nonlinear wavelet approach to nonparametric regression and introduces several recent thresholding techniques that can be formulated as penalized least squares problems. We also discuss some block-wise thresholding procedures that have been shown to enjoy a high level of adaptation for wavelet series estimates and end with a short discussion on block thresholding methods for density estimation. Few wavelet based applications to complex problems are given in Section 4. Whenever necessary, the practical performance of the methods that are discussed is examined by appropriate simulations.

## 2. A short background on wavelets

In this section we give a brief overview of some relevant material on the wavelet series expansion and a fast wavelet transform that we need further.

### 2.1. The wavelet series expansion

The term *wavelets* is used to refer to a set of orthonormal basis functions generated by dilation and translation of a compactly supported *scaling function* (or *father wavelet*), $\phi$, and a *mother wavelet*, $\psi$, associated with an $r$-regular multiresolution analysis of $L^2(\mathbb{R})$. A variety of different *wavelet families* now exist that combine compact support with various degrees of smoothness and numbers of vanishing moments (see, Daubechies (1992)), and these are now the most intensively used wavelet families in practical applications in statistics. Hence, many types of functions encountered in practice can be sparsely (i.e. parsimoniously) and uniquely represented in terms of a wavelet series. Wavelet bases are therefore not only useful by virtue of their special structure, but they may also be (and have been!) applied in a wide variety of contexts.

For simplicity in exposition, we shall assume that we are working with periodized wavelet bases on $[0, 1]$ (see, for example, Mallat (1999), Section 7.5.1), letting

$$\phi^{\mathrm{p}}_{jk}(t) = \sum_{l \in \mathbb{Z}} \phi_{jk}(t-l) \quad \text{and} \quad \psi^{\mathrm{p}}_{jk}(t) = \sum_{l \in \mathbb{Z}} \psi_{jk}(t-l), \quad \text{for} \quad t \in [0, 1],$$

where

$$\phi_{jk}(t) = 2^{j/2}\phi(2^j t - k), \quad \psi_{jk}(t) = 2^{j/2}\psi(2^j t - k).$$

For any $j_0 \geq 0$, the collection $\{\phi^{\mathrm{p}}_{j_0 k}, \ k = 0, 1, \ldots, 2^{j_0} - 1; \ \psi^{\mathrm{p}}_{j_0 k}, \ j \geq j_0 \geq 0, \ k = 0, 1, \ldots, 2^j - 1\}$ is then an orthonormal basis of $L^2([0, 1])$. The superscript "p" will be suppressed from the notation for convenience.

The idea underlying such an approach is to express any function $g \in L^2([0, 1])$ in the form

$$g(t) = \sum_{k=0}^{2^{j_0}-1} \alpha_{j_0 k} \phi_{j_0 k}(t) + \sum_{j=j_0}^{\infty} \sum_{k=0}^{2^j-1} \beta_{jk} \psi_{jk}(t), \quad j_0 \geq 0, \quad t \in [0, 1],$$

where

$$\alpha_{j_0 k} = \langle g, \phi_{j_0 k} \rangle = \int_0^1 g(t) \phi_{j_0 k}(t) \, dt, \quad j_0 \geq 0, \quad k = 0, 1, \ldots, 2^{j_0} - 1$$

and

$$\beta_{jk} = \langle g, \psi_{jk} \rangle = \int_0^1 g(t) \psi_{jk}(t) \, dt, \quad j \geq j_0 \geq 0, \quad k = 0, 1, \ldots, 2^j - 1.$$

An usual assumption underlying the use of periodic wavelets is that the function to be expanded is assumed to be periodic. However, such an assumption is not always realistic and periodic wavelets exhibit a poor behaviour near the boundaries (they create high amplitude wavelet coefficients in the neighborhood of the boundaries when the analysed function is not periodic). However, periodic wavelets are commonly used because the numerical implementation is particular simple. While, as Johnstone (1994) has pointed out, this computational simplification affects only a fixed number of wavelet coefficients at each resolution level, we will also present later on an effective method, developed recently by Oh and Lee (2005), combining wavelet decompositions with local polynomial regression, for correcting the boundary bias introduced by the inappropriateness of the periodic assumption.

### 2.2. Function spaces and wavelets

The (inhomogeneous) Besov spaces on the unit interval, $B^s_{\rho_1,\rho_2}[0,1]$, consist of functions that have a specific degree of smoothness in their derivatives. More specifically, let the $r$th difference of a function $f$ be

$$\Delta^{(r)}_h f(t) = \sum_{k=0}^r \binom{r}{k}(-1)^k f(t+kh),$$

and let the $r$th modulus of smoothness of $f \in L^{\rho_1}[0,1]$ be

$$\nu_{r,\rho_1}(f;t) = \sup_{h \leq t}(||\Delta^{(r)}_h f||_{L^{\rho_1}[0,1-rh]}).$$

Then the Besov seminorm of index $(s,\rho_1,\rho_2)$ is defined for $r > s$, where $1 \leq \rho_1, \rho_2 \leq \infty$, by

$$|f|_{B^s_{\rho_1,\rho_2}} = \left[\int_0^1 \left\{\frac{\nu_{r,\rho_1}(f;h)}{h^s}\right\}^{\rho_2} \frac{dh}{h}\right]^{1/\rho_2}, \qquad \text{if} \qquad 1 \leq \rho_2 < \infty,$$

and by

$$|f|_{B^s_{\rho_1,\infty}} = \sup_{0<h<1} \left\{\frac{\nu_{r,\rho_1}(f;h)}{h^s}\right\}.$$

The Besov norm is then defined as

$$||f||_{B^s_{\rho_1,\rho_2}} = ||f||_{L^{\rho_1}} + |f|_{B^s_{\rho_1,\rho_2}}$$

and the Besov space on $[0,1]$, $B^s_{\rho_1,\rho_2}[0,1]$, is the class of functions $f : [0,1] \to \mathbb{R}$ satisfying $f \in L^{\rho_1}[0,1]$ and $|f|_{B^s_{\rho_1,\rho_2}} < \infty$, i.e. satisfying $||f||_{B^s_{\rho_1,\rho_2}} < \infty$. The parameter $\rho_1$ can be viewed as a degree of function's inhomogeneity while $s$ is a measure of its smoothness. Roughly speaking, the (not necessarily integer) parameter $s$ indicates the number of function's derivatives, where their existence is required in an $L^{\rho_1}$-sense; the additional parameter $\rho_2$ is secondary in its role, allowing for additional fine tuning of the definition of the space.

The Besov classes include, in particular, the well-known Hilbert-Sobolev ($H^s_2[0,1]$, $s = 1,2,\ldots$) and Hölder ($C^s[0,1]$, $s > 0$) spaces of smooth functions ($B^s_{2,2}[0,1]$ and $B^s_{\infty,\infty}[0,1]$ respectively), but in addition less-traditional spaces, like the space of bounded-variation, sandwiched between $B^1_{1,1}[0,1]$ and $B^1_{1,\infty}[0,1]$. The latter functions are of statistical interest because they allow for better models of spatial inhomogeneity (see, e.g., Meyer (1992)).

The Besov norm for the function $f$ is related to a sequence space norm on the wavelet coefficients of the function. Confining attention to the resolution and spatial indices $j \geq j_0$ and $k = 0,1,\ldots,2^j-1$ respectively, and denoting by $s' = s + 1/2 - 1/\rho_1$, the sequence space norm is given by

$$
\begin{aligned}
||\boldsymbol{\theta}||_{b^s_{\rho_1,\rho_2}} &= ||\boldsymbol{\alpha}_{j_0}||_{\rho_1} + \left\{\sum_{j=j_0}^\infty 2^{js'\rho_2}||\boldsymbol{\beta}_j||^{\rho_2}_{\rho_1}\right\}^{1/\rho_2}, \quad \text{if} \quad 1 \leq \rho_2 < \infty, \\
||\boldsymbol{\theta}||_{b^s_{\rho_1,\infty}} &= ||\boldsymbol{\alpha}_{j_0}||_{\rho_1} + \sup_{j \geq j_0}\left\{2^{js'}||\boldsymbol{\beta}_j||_{\rho_1}\right\},
\end{aligned}
$$

where

$$||\boldsymbol{\alpha}_{j_0}||_{\rho_1}^{\rho_1} = \sum_{k=0}^{2^{j_0}-1} |\alpha_{j_0 k}|^{\rho_1} \quad \text{and} \quad ||\boldsymbol{\beta}_j||_{\rho_1}^{\rho_1} = \sum_{k=0}^{2^j-1} |\beta_{jk}|^{\rho_1}.$$

If the mother wavelet $\psi$ is of regularity $r > 0$, it can be shown that the corresponding orthonormal periodic wavelet basis defined in Section 2.1 is an unconditional basis for the Besov spaces $B_{\rho_1,\rho_2}^s[0,1]$ for $0 < s < r$, $1 \leq \rho_1, \rho_2 \leq \infty$. In other words, we have

$$K_1||f||_{B_{\rho_1,\rho_2}^s} \leq ||\boldsymbol{\theta}||_{b_{\rho_1,\rho_2}^s} \leq K_2||f||_{B_{\rho_1,\rho_2}^s},$$

where $K_1$ and $K_2$ are constants, not depending on $f$. Therefore the Besov norm of the function $f$ is equivalent to the corresponding sequence space norm defined above; this allows one to characterize Besov spaces in terms of wavelet coefficients (see, e.g., Meyer (1992); Donoho and Johnstone (1998)). For a more detailed study on (inhomogeneous) Besov spaces we refer to Meyer (1992).

### 2.3. The discrete wavelet transform

In statistical settings we are more usually concerned with discretely sampled, rather than continuous, functions. It is then the wavelet analogy to the discrete Fourier transform which is of primary interest and this is referred to as the discrete wavelet transform (DWT). Given a vector of function values $\mathbf{g} = (g(t_1), ..., g(t_n))'$ at equally spaced points $t_i$, the discrete wavelet transform of $\mathbf{g}$ is given by

$$\mathbf{d} = W\mathbf{g},$$

where $\mathbf{d}$ is an $n \times 1$ vector comprising both discrete scaling coefficients, $c_{j_0 k}$, and discrete wavelet coefficients, $d_{jk}$, and $W$ is an orthogonal $n \times n$ matrix associated with the orthonormal wavelet basis chosen. The $c_{j_0 k}$ and $d_{jk}$ are related to their continuous counterparts $\alpha_{j_0 k}$ and $\beta_{jk}$ (with an approximation error of order $n^{-1}$) via the relationships

$$c_{j_0 k} \approx \sqrt{n} \, \alpha_{j_0 k} \quad \text{and} \quad d_{jk} \approx \sqrt{n} \, \beta_{jk}.$$

The factor $\sqrt{n}$ arises because of the difference between the continuous and discrete orthonormality conditions. This root factor is unfortunate but both the definition of the DWT and the wavelet coefficients are now fixed by convention, hence the different notation used to distinguish between the discrete wavelet coefficients and their continuous counterpart. Note that, because of orthogonality of $W$, the inverse DWT (IDWT) is simply given by

$$\mathbf{g} = W^T\mathbf{d},$$

where $W^T$ denotes the transpose of $W$.

If $n = 2^J$ for some positive integer $J$, the DWT and IDWT may be performed through a computationally fast algorithm developed by Mallat (1999) that requires only order $n$ operations. In this case, for a given $j_0$ and under periodic boundary conditions, the DWT of $\mathbf{g}$ results in an $n$-dimensional vector $\mathbf{d}$ comprising both discrete scaling coefficients $c_{j_0 k}$, $k = 0, ..., 2^{j_0} - 1$ and discrete wavelet coefficients $d_{jk}$, $j = j_0, ..., J - 1$; $k = 0, ..., 2^j - 1$.

We do not provide technical details here of the order $n$ DWT algorithm mentioned above. Essentially the algorithm is a fast hierarchical scheme for deriving the required inner products which at each step involves the action of low and high pass filters, followed by a decimation (selection of every even member of a sequence). The IDWT may be similarly obtained in terms of related filtering operations. For excellent accounts of the DWT and IDWT in terms of filter operators we refer to Nason and Silverman (1995), Strang and Nguyen (1996), or Burrus et al. (1998).

## 3. Denoising by wavelet thresholding

The problem of estimating a signal that is corrupted by additive noise is a standard problem in statistics and signal processing. It can be described as follows. Consider the standard univariate nonparametric regression setting

$$y_i = g(t_i) + \sigma \ \epsilon_i, \quad i = 1, \ldots, n, \tag{3.1}$$

where $\epsilon_i$ are independent $N(0,1)$ random variables and the noise level $\sigma$ may be known or unknown. We suppose, without loss of generality, that $t_i$ are within the unit interval $[0,1]$. The goal is to recover the underlying function $g$ from the noisy data, $\mathbf{y} = (y_1, \ldots, y_n)^T$, without assuming any particular parametric structure for $g$.

One of the basic approaches to handle this regression problem is to consider the unknown function $g$ expanded as a generalised Fourier series and then to estimate the generalised Fourier coefficients from the data. The original (nonparametric) problem is thus transformed to a parametric one, although the potential number of parameters is infinite. An appropriate choice of basis for the expansion is therefore a key point in relation to the efficiency of such an approach. A 'good' basis should be parsimonious in the sense that a large set of possible response functions can be approximated well by only few terms of the generalized Fourier expansion employed. Wavelet series allow a parsimonious expansion for a wide variety of functions, including inhomogeneous cases. It is therefore natural to consider applying the generalized Fourier series approach using a wavelet series.

In what follows we assume that the sample points are equally spaced, i.e. $t_i = i/n$, and that the sample size $n$ is a power of two: $n = 2^J$ for some positive integer $J$. These assumptions allow us to perform both the DWT and the IWDT using Mallat's fast algorithm. Note, however, that for non-equispaced or random designs, or sample sizes which are not a power of two, or data contaminated with correlated noise, modifications are needed to the standard wavelet-based estimation procedures that will be discussed in subsection 3.4.

Due to the orthogonality of the matrix $W$, the DWT of white noise is also an array of independent $N(0, 1)$ random variables, so from (3.1) it follows that

$$\hat{c}_{j_0 k} = c_{j_0 k} + \sigma \, \epsilon_{jk}, \quad k = 0, 1, \ldots, 2^{j_0} - 1, \tag{3.2}$$

$$\hat{d}_{jk} = d_{jk} + \sigma \, \epsilon_{jk}, \quad j = j_0, \ldots, J - 1, \quad k = 0, \ldots, 2^j - 1, \tag{3.3}$$

where $\hat{c}_{j_0 k}$ and $\hat{d}_{jk}$ are respectively the *empirical scaling* and the *empirical wavelet* coefficients of the noisy data **y**, and $\epsilon_{jk}$ are independent $N(0, 1)$ random variables.

The sparseness of the wavelet expansion makes it reasonable to assume that essentially only a few 'large' $d_{jk}$ contain information about the underlying function $g$, while 'small' $d_{jk}$ can be attributed to the noise which uniformly contaminates all wavelet coefficients. Thus, simple denoising algorithms that use the wavelet transform consist of three steps:

1) Calculate the wavelet transform of the noisy signal.
2) Modify the noisy wavelet coefficients according to some rule.
3) Compute the inverse transform using the modified coefficients.

Traditionally, for the second step of the above approach there are two kinds of denoising methods; namely, linear and nonlinear techniques. A wavelet based linear approach, extending simply spline smoothing estimation methods as described by Wahba (1990), is the one suggested by Antoniadis (1996) and independently by Amato and Vuza (1997). This method is appropriate for estimating relatively regular functions. Assuming that the smoothness index $s$ of the function $g$ to be recovered is known, the resulting estimator is obtained by estimating the scaling coefficients $c_{j_0 k}$ by their empirical counterparts $\hat{c}_{j_0 k}$ and by estimating the wavelet coefficients $d_{jk}$ via a linear shrinkage

$$\tilde{d}_{jk} = \frac{\hat{d}_{jk}}{1 + \lambda 2^{2js}},$$

where $\lambda > 0$ is a smoothing parameter. The parameter $\lambda$ is chosen by cross-validation in Amato and Vuza (1997), while the choice of $\lambda$ in Antoniadis (1996) is based on risk minimization and depends on a preliminary consistent estimator of the noise level $\sigma$. While simple and cheap to implement, the above linear method *is not* designed to handle spatially inhomogeneous functions with *low* regularity. For such functions one usually relies upon nonlinear thresholding or shrinkage methods. One of the earliest papers in the field of wavelet denoising may be that of Weaver et al. (1991), proposing a hard-thresholding scheme for filtering noise from magnetic resonance images. While Weaver et al. (1991) demonstrated the advantages of the wavelet thresholding scheme mainly based on experimental results, the first thorough mathematical treatment of wavelet shrinkage and wavelet thresholding was done by Donoho *et al.* in a series of technical reports in the early 1990's and published in Donoho and Johnstone (1994), Donoho (1995), Donoho et al. (1995) and Donoho and Johnstone (1998). Donoho and his coworkers analyzed wavelet thresholding and shrinkage methods in the context of minimax estimation and showed, that wavelet shrinkage generates asymptotically optimal estimates for noisy data that outperform any linear estimator.

Mathematically wavelet coefficients are estimated using either the *hard* or *soft* thresholding rule given respectively by

$$\delta_\lambda^{\mathrm{H}}(\hat{d}_{jk}) = \begin{cases} 0 & \text{if } |\hat{d}_{jk}| \leq \lambda \\ \hat{d}_{jk} & \text{if } |\hat{d}_{jk}| > \lambda \end{cases} \tag{3.4}$$

and

$$\delta_\lambda^{\mathrm{S}}(\hat{d}_{jk}) = \begin{cases} 0 & \text{if } |\hat{d}_{jk}| \leq \lambda \\ \hat{d}_{jk} - \lambda & \text{if } \hat{d}_{jk} > \lambda \\ \hat{d}_{jk} + \lambda & \text{if } \hat{d}_{jk} < -\lambda. \end{cases} \tag{3.5}$$

Thresholding allows the data itself to decide which wavelet coefficients are significant; hard thresholding (a discontinuous function) is a 'keep' or 'kill' rule, while soft thresholding (a continuous function) is a 'shrink' or 'kill' rule. Beside these two possibilities there are many others (semi-soft shrinkage, firm shrinkage, . . . ) and as long as the shrinkage function preserves the sign $(\mathrm{sign}(\delta_\lambda(x)) = \mathrm{sign}(x))$ and shrinks the magnitude $(|\delta_\lambda(x)| \leq |x|)$ one can expect a denoising effect.

Gao and Bruce (1997) and Marron et al. (1998) have shown that simple threshold values with hard thresholding results in larger variance in the function estimate, while the same threshold values with soft thresholding shift the estimated coefficients by an amount of $\lambda$ even when $|\hat{d}_{jk}|$ stand way out of noise level, creating unnecessary bias when the true coefficients are large. Also, due to its discontinuity, hard thresholding can be unstable – that is, sensitive to small changes in the data.

To remedy the drawbacks of both hard and soft thresholding rules, Gao and Bruce (1997) considered the *firm* threshold thresholding

$$\delta_{\lambda_1,\lambda_2}^{\mathrm{F}}(\hat{d}_{jk}) = \begin{cases} 0 & \text{if } |\hat{d}_{jk}| \leq \lambda_1 \\ \mathrm{sign}(\hat{d}_{jk})\frac{\lambda_2(|\hat{d}_{jk}|-\lambda_1)}{\lambda_2 - \lambda_1} & \text{if } \lambda_1 < |\hat{d}_{jk}| \leq \lambda_2 \\ \hat{d}_{jk} & \text{if } |\hat{d}_{jk}| > \lambda_2 \end{cases} \tag{3.6}$$

which is a "shrink" or "kill" rule (a continuous function).

The resulting wavelet thresholding estimators offer, in small samples, advantages over both hard thresholding (generally smaller mean squared error and less sensitivity to small perturbations in the data) and soft thresholding (generally smaller bias and overall mean squared error) rules. For values of $|\hat{d}_{jk}|$ near the lower threshold $\lambda_1$, $\delta_{\lambda_1,\lambda_2}^{\mathrm{F}}(\hat{d}_{jk})$ behaves like $\delta_{\lambda_1}^{\mathrm{S}}(\hat{d}_{jk})$. For values of $|\hat{d}_{jk}|$ above the upper threshold $\lambda_2$, $\delta_{\lambda_1,\lambda_2}^{\mathrm{F}}(\hat{d}_{jk})$ behaves like $\delta_{\lambda_2}^{\mathrm{H}}(\hat{d}_{jk})$. Note that the hard thresholding and soft thresholding rules are limiting cases of (3.6) with $\lambda_1 = \lambda_2$ and $\lambda_2 = \infty$ respectively.

Note that firm thresholding has a drawback in that it requires two threshold values (one for 'keep' or 'shrink' and another for 'shrink' or 'kill'), thus making the estimation procedure for the threshold values more computationally expensive. To overcome this drawback, Gao (1998) considered the *nonnegative garrote*

thresholding

$$\delta_\lambda^{\mathrm{G}}(\hat{d}_{jk}) = \begin{cases} 0 & \text{if } |\hat{d}_{jk}| \le \lambda \\ \hat{d}_{jk} - \frac{\lambda^2}{\hat{d}_{jk}} & \text{if } |\hat{d}_{jk}| > \lambda \end{cases} \tag{3.7}$$

which also is a "shrink" or "kill" rule (a continuous function). The resulting wavelet thresholding estimators offer, in small samples, advantages over both hard thresholding and soft thresholding rules that is comparable to the firm thresholding rule, while the latter requires two threshold values.

In the same spirit to that in Gao (1998), Antoniadis and Fan (2001) suggested the *SCAD* thresholding rule

$$\delta_\lambda^{\mathrm{SCAD}}(\hat{d}_{jk}) = \begin{cases} \operatorname{sign}(\hat{d}_{jk}) \max(0, |\hat{d}_{jk}| - \lambda) & \text{if } |\hat{d}_{jk}| \le 2\lambda \\ \frac{(a-1)\hat{d}_{jk} - a\lambda \operatorname{sign}(\hat{d}_{jk})}{a-2} & \text{if } 2\lambda < |\hat{d}_{jk}| \le a\lambda \\ \hat{d}_{jk} & \text{if } |\hat{d}_{jk}| > a\lambda \end{cases} \tag{3.8}$$

which is a "shrink" or "kill" rule (a piecewise linear function). It does not over penalize large values of $|\hat{d}_{jk}|$ and hence does not create excessive bias when the wavelet coefficients are large. Antoniadis and Fan (2001), based on a Bayesian argument, have recommended to use the value of $a = 3.7$.

The interesting thing about wavelet shrinkage is, that it can be interpreted in very different ways, less known in the statistics literature, allowing a better understanding of wavelet shrinkage and providing an unified framework for many seemingly different thresholding rules in nonparametric function estimation.

### 3.1. Wavelet shrinkage and nonlinear diffusion

Nonlinear diffusion filtering and wavelet shrinkage are two methods that serve the same purpose, namely discontinuity-preserving denoising. In this subsection we give a survey on relations between both paradigms when fully discrete versions of nonlinear diffusion filters with an explicit time discretization are considered. In particular we present the results of Mrázek et al. (2003) connecting a shift-invariant Haar wavelet shrinkage and the diffusivity of a nonlinear diffusion filter. This allows to present corresponding diffusivity functions to some known and widely used shrinkage functions or new shrinkage functions with competitive performance which are induced by famous diffusivities. Due to the lack of space we can only present the main ideas and refer the reader to the paper of Mrázek et al. (2003) for more details. Before proceeding, we would like first to recall some facts about translation-invariant denoising.

One drawback of the discrete wavelet transform is that the coefficients of the discretized signal are not circularly shift equivariant, so that circularly shifting the observed series by some amount will not circularly shift the discrete wavelet transform coefficients by the same amount. This seriously degrades the quality of the denoising achieved. To try to alleviate this problem Coifman and Donoho (1995) introduced the technique of 'cycle spinning'. The idea of denoising via cycle spinning is to apply denoising not only to $\mathbf{y}$, but also to all possible unique circularly shifted versions of **y**, and to average the results. As pointed out by Percival and Walden (2000) (see, p. 429) another way to perform the translation invariant shrinkage is by applying standard thresholding to the wavelet coefficients of the maximal overlap discrete wavelet transform, a transform we more briefly refer to as the stationary wavelet transform and we refer to the above monograph for details. We are now able to introduce and analyze a general connection between translation invariant Haar wavelet shrinkage and a discretized version of a nonlinear diffusion.

The scaling and wavelet filters $h$ and $\tilde{h}$ corresponding to the Haar wavelet transform are

$$h = \frac{1}{\sqrt{2}}(\dots, 0, 1, 1, 0, \dots) \quad \tilde{h} = \frac{1}{\sqrt{2}}(\dots, 0, -1, 1, 0, \dots).$$

Given a discrete signal $f = (f_k)_{k \in \mathbb{Z}}$, it is easy to see that a shift-invariant soft wavelet shrinkage of $f$ on a single level decomposition with the Haar wavelet creates a filtered signal $u = (u_k)_{k \in \mathbb{Z}}$ given by

$$u_k = \frac{1}{4}(f_{k-1} + 2f_k + f_{k+1}) + \frac{1}{2\sqrt{2}}\left(-\delta_\lambda^{\mathrm{S}}\left(\frac{f_{k+1} - f_k}{\sqrt{2}}\right) + \delta_\lambda^{\mathrm{S}}\left(\frac{f_k - f_{k-1}}{\sqrt{2}}\right)\right),$$

where $\delta_\lambda^{\mathrm{S}}$ denotes the soft shrinkage operator with threshold $\lambda$. Because the filters of the Haar wavelet are simple difference filters (a finite difference approximation of derivatives) the above shrinkage rule looks a little like a discretized version of a differential equation. Indeed, rewriting the above equation as

$$\begin{aligned}
u_k &= f_k + \frac{f_{k+1} - f_k}{4} - \frac{f_k - f_{k-1}}{4} \\
&\quad + \frac{1}{2\sqrt{2}}\left(-\delta_\lambda^{\mathrm{S}}\left(\frac{f_{k+1} - f_k}{\sqrt{2}}\right) + \delta_\lambda^{\mathrm{S}}\left(\frac{f_k - f_{k-1}}{\sqrt{2}}\right)\right) \\
&= f_k + \left(\frac{(f_{k+1} - f_k)}{4} - \frac{1}{2\sqrt{2}}\delta_\lambda^{\mathrm{S}}\left(\frac{f_{k+1} - f_k}{\sqrt{2}}\right)\right) \\
&\quad - \left(\frac{(f_k - f_{k-1})}{4} - \frac{1}{2\sqrt{2}}\delta_\lambda^{\mathrm{S}}\left(\frac{f_k - f_{k-1}}{\sqrt{2}}\right)\right),
\end{aligned}$$

we obtain

$$\frac{u_k - f_k}{\Delta t} = (f_{k+1} - f_k)g(|f_{k+1} - f_k|) - (f_k - f_{k-1})g(|f_k - f_{k-1}|), \quad (3.9)$$

with a function $g$ and a time step size $\Delta t$ defined by

$$\Delta t \, g(|s|) = \frac{1}{4} - \frac{1}{2\sqrt{2}|s|}\delta_\lambda^{\mathrm{S}}\left(\frac{|s|}{\sqrt{2}}\right).$$

Eq. (3.9) appears as a first iteration of an explicit (Euler forward) scheme for a nonlinear diffusion filter with initial state $f$, time step size $\Delta t$ and spatial step size 1, and therefore the shrinkage rule corresponds to a discretization of the following differential equation

$$\partial_t u = \partial_x \left((\partial_x u)g(|\partial_x u|)\right), \quad (3.10)$$

with initial condition $u(0) = f$. This equation is a 1-D variant of the Perona-Malik diffusion equation well known in image processing, and the function $g$ is called the diffusivity. The basic idea behind nonlinear diffusion filtering in the 1-D case (see Droske and Rumpf (2004)) is to obtain a family $u(x, t)$ of filtered versions of a continuous signal $f$ as the solution of the diffusion process stated in Eq. (3.10) with f as initial condition, $u(x, 0) = f(x)$ and reflecting boundary conditions. The diffusivity $g$ controls the speed of diffusion depending on the magnitude of the gradient. Usually, $g$ is chosen such that it is equal to one for small magnitudes of the gradient and goes down to zero for large gradients. Hence the diffusion stops at positions where the gradient is large. These areas are considered as singularities of the signal. Since the Perona-Malik equation is nonlinear the existence of a solution is not obvious.

We can now give a proposition which relates some properties of shrinkage functions and diffusivities and whose proof is an easy consequence of the relation between $g$ and $\delta_\lambda$. A detailed analysis of this connection in terms of extremum principles, monotonicity preservation and sign stability can be found in Mrázek et al. (2003) (see also Lorenz (2006)). We formulate this relations for the case $\Delta t = 1/4$ which is a common choice and widely used for the Perona-Malik equation. This choice makes the relations more clear and relates some nice properties of the shrinkage function to other nice properties of the diffusivity.

**Propostion 3.1.** *Let $\Delta t = 1/4$. Then the diffusivity and the shrinkage function are related through*

$$g(|x|) = 1 - \frac{\sqrt{2}}{|x|} \delta_\lambda \left( \frac{|x|}{\sqrt{2}} \right). \tag{3.11}$$

*The following properties hold:*

1. *If $\delta_\lambda$ performs shrinkage then the diffusion is always forward, i. e.*

$$\delta_\lambda(|x|) \leq |x| \Leftrightarrow g(x) \geq 0.$$

2. *If $\delta_\lambda$ is differentiable at 0 then, as $x \to 0$,*

$$g(x) \to 1 \Leftrightarrow \delta_\lambda(0) = 0 \text{ and } \delta_\lambda'(0) = 0.$$

3. *If the diffusion stops for large gradients the shrinkage function has linear growth at infinity, i. e.*

$$g(x) \to 0, \text{ as } x \to \infty \Leftrightarrow \frac{\delta_\lambda(x)}{x} \to 1, \text{ as } x \to \infty.$$

Some examples will make the correspondence more clear. As suggested by Proposition 3.1 we choose $\Delta t = 1/4$ and derive the corresponding diffusivities by plug in the specific shrinkage function into (3.11).

**Linear shrinkage** A linear shrinkage rule, producing linear wavelet denoising is given by $\delta_\lambda(x) = \frac{x}{1+\lambda}$. The corresponding diffusivity is constant $g(|x|) = \frac{\lambda}{(1+\lambda)}$, and the diffusion is linear.

**Soft shrinkage** The soft shrinkage function $\delta_\lambda(x) = \text{sign}(x)(|x| - \lambda)_+$ gives $g(|x|) = \left(1 - \frac{(|x| - \sqrt{2}\lambda)_+}{|x|}\right)$, which is a stabilized total variation diffusivity (see Steidl and Weickert (2002)).

**Hard shrinkage** The hard shrinkage function $\delta_\lambda(x) = x(1 - I_{\{|x| \le \lambda\}}(x))$ leads to $g(|x|) = I_{\{|x| \le \sqrt{2}\lambda\}}(|x|)$ which is a piecewise linear diffusion that degenerates for large gradients.

**Garrote shrinkage** The nonnegative garrote shrinkage $\delta_\lambda(x) = \left(x - \frac{\lambda^2}{x}\right)\left(1 - I_{\{|x| \le \lambda\}}(x)\right)$ leads to a stabilized unbounded BFB diffusivity given by $g(|x|) = I_{\{|x| \le \sqrt{2}\lambda\}}(|x|) + \frac{2\lambda^2}{x^2}I_{\{|x| > \sqrt{2}\lambda\}}(|x|)$.

**Firm shrinkage** Firm shrinkage defined by eq. (3.6) yields a diffusivity that degenerates to 0 for sufficiently large gradients:

$$g(|x|) = \begin{cases} 1 & \text{if } |x| \le \sqrt{2}\lambda_1 \\ \frac{\lambda_1}{(\lambda_2 - \lambda_1)}\left(\frac{\sqrt{2}\lambda_2}{|x|} - 1\right) & \text{if } \sqrt{2}\lambda_1 < |x| \le \sqrt{2}\lambda_2 \\ 0 & \text{if } |x| > \sqrt{2}\lambda_2 \end{cases}.$$

**SCAD shrinkage** SCAD shrinkage defined by eq. (3.8) gives also a diffusivity that degenerates to 0:

$$g(|x|) = \begin{cases} 1 & \text{if } |x| \le \sqrt{2}\lambda \\ \frac{\sqrt{2}\lambda}{|x|} & \text{if } \sqrt{2}\lambda < |x| \le 2\sqrt{2}\lambda \\ \frac{a\sqrt{2}\lambda}{(a-2)|x|} - \frac{1}{a-2} & \text{if } 2\sqrt{2}\lambda < |x| \le a\sqrt{2}\lambda \\ 0 & \text{if } |x| > a\sqrt{2}\lambda \end{cases}.$$

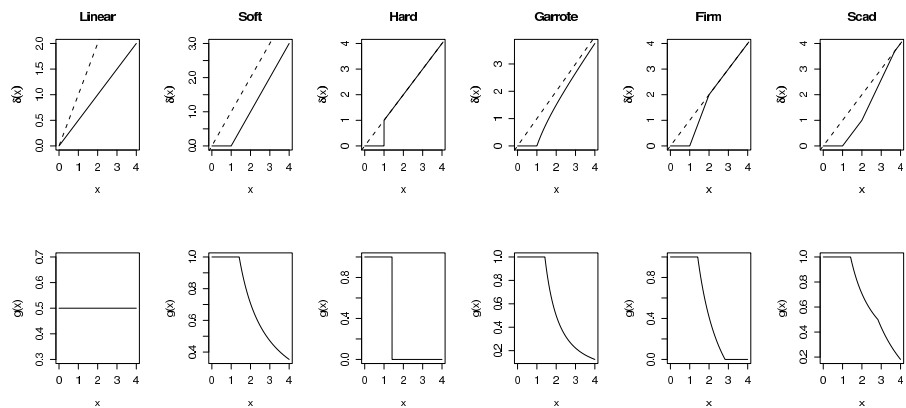

Fig 1. *Shrinkage functions (top) and corresponding diffusivities (bottom). The functions are plotted for $\lambda = 1$, $\lambda_1 = 1$, $\lambda_2 = 2$ (Firm) and $a = 3.7$ (Scad). The dashed line represents the diagonal.*

All these example are depicted in Figure 1. The other way round one can ask, how the shrinkage functions for famous diffusivities look like. The function $\delta_\lambda$ expressed in terms of $g$ looks like $\delta_\lambda(|x|) = |x|(1 - g(\sqrt{2}|x|))$ and the dependence of the shrinkage function on the threshold parameter $\lambda$ is naturally fulfilled because usually diffusivities involve a parameter too. Using this remark we obtain the following new shrinkage functions:

**Charbonnier diffusivity** The Charbonnier diffusivity (Charbonnier et al. (1994)) is given by $g(|x|) = \left(1 + \frac{x^2}{\lambda^2}\right)^{-1/2}$ and corresponds to the shrinkage function $\delta_\lambda(x) = x\left(1 - \sqrt{\frac{\lambda^2}{\lambda^2 + 2x^2}}\right)$.

**Perona-Malik diffusivity** The Perona-Malik diffusivity (Perona and Malik (1990)) is defined by $g(|x|) = \left(1 + \frac{x^2}{\lambda^2}\right)^{-1}$ and lead to the shrinkage function $\delta_\lambda(x) = \frac{2x^3}{2x^2 + \lambda^2}$.

**Weickert diffusivity** Weickert (1998) introduced the following diffusivity $g(|x|) = I_{\{|x|>0\}}(x)\left(1 - \exp\left(-\frac{3.31488}{(|x|/\lambda)^8}\right)\right)$ which leads to the shrinkage function $\delta_\lambda(x) = x\exp\left(-\frac{0.20718\lambda^8}{x^8}\right)$.

**Tukey diffusivity** Tukey's diffusivity, defined by Black et al. (1998) as $g(|x|) = (1 - (x/\lambda)^2)^2 I_{\{|x|\le\lambda\}}(|x|)$ leads to the shrinkage function

$$\delta_\lambda(x) = \begin{cases} \frac{4x^3}{\lambda^2} - \frac{4x^5}{\lambda^4} & \text{if } |x| \le \lambda/\sqrt{2} \\ x & \text{if } |x| > \lambda/\sqrt{2} \end{cases}.$$

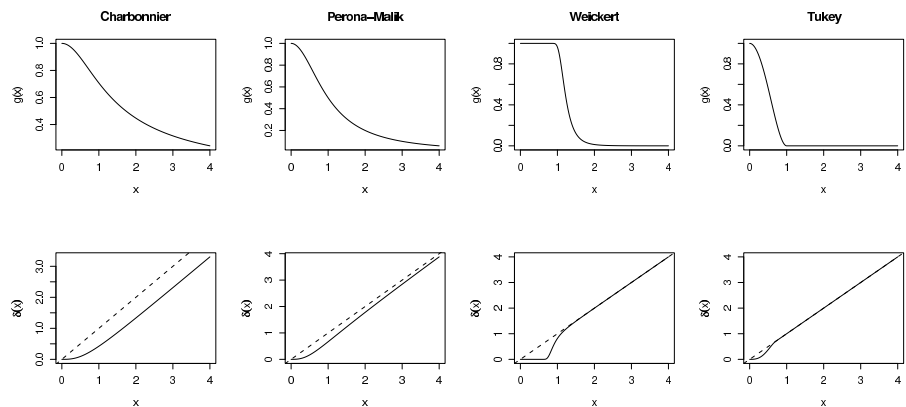

Fig 2. *"Classical" diffusivites (top) and corresponding shrinkage functions.*

Figure 2 illustrates these diffusivities and their corresponding shrinkage functions. Having developed the connection between diffusivities and shrinkage functions, we will further exploit them in the subsection that follows in order to show that they can all be interpreted as cases of a broad class of penalized least squares estimators. This unified treatment and the general results of Antoniadis and Fan (2001) on penalized wavelet estimators allow a systematic derivation of oracle inequalities and minimax properties for a large class of wavelet estimators.

### 3.2. Penalized least-squares wavelet estimators

The various thresholding rules presented in the previous subsection play no monopoly in choosing an ideal wavelet sub-basis to efficiently estimate an unknown function observed with noise. It turns out that the corresponding nonlinear estimators can be seen as optimal estimators in a regularization setting for specific penalty functions, i.e. most of the shrinkage rules can be linked to a regularization process under a corresponding and reasonable penalty function. Exploring the nature of these penalties and using results from Antoniadis and Fan (2001), it is then easy to show that the corresponding thresholding estimators have good sampling properties and are adaptively minimax. It is the aim of this subsection to integrate the diverse shrinkage rules from a regularization point of view.

When estimating a signal that is corrupted by additive noise by wavelet based methods, the traditional regularization problem can be formulated in the wavelet domain by finding the minimum in $\boldsymbol{\theta}$ of the penalized least-squares functional $\ell(\boldsymbol{\theta})$ defined by

$$\ell(\boldsymbol{\theta}) = \|W\mathbf{y} - \boldsymbol{\theta}\|_n^2 + 2\lambda \sum_{i>i_0} p(|\theta_i|), \qquad (3.12)$$

where $\boldsymbol{\theta}$ is the vector of the wavelet coefficients of the unknown regression function $g$ and $p$ is a given penalty function, while $i_0$ is a given integer corresponding to penalizing wavelet coefficients above certain resolution level $j_0$. Here to facilitate the presentation we changed the notation $d_{j,k}$ from a double array sequence into a single array sequence $\theta_i$. Similarly, denote by $\mathbf{z}$ the vector of empirical wavelet coefficients $W\mathbf{y}$. With a choice of an additive penalty $\sum_{i>i_0} p(|\theta_i|)$, the minimization problem becomes separable. Minimizing (3.12) is equivalent to minimizing

$$\ell(\theta_i) = \|z_i - \theta_i\|^2 + 2\lambda p(|\theta_i|), \qquad (3.13)$$

for each coordinate $i$. The resulting penalized least-squares estimator is in this case separable, that is the estimate of any coordinate $\theta_i$ depends solely on the empirical wavelet coefficient $z_i$. While separable estimators have their drawbacks, this is the case that we address here. To reduce abuse of notation, and because $p(|\theta|)$ is allowed to depend on $\lambda$, we will use $p_\lambda$ to denote the penalty function $\lambda p$ in the following discussion.

The performance of the resulting wavelet estimator depends on the penalty and the regularization parameter $\lambda$. To select a good penalty function, Antoniadis and Fan (2001) and Fan and Li (2001) proposed three principles that a good penalty function should satisfy: unbiasedness, in which there is no over-penalization of large coefficients to avoid unnecessary modeling biases; sparsity, as the resulting penalized least-squares estimators should follow a thresholding rule such that insignificant coefficients should be set to zero to reduce model complexity; and continuity to avoid instability and large variability in model prediction. The interested reader is referred to Theorem 1 of Antoniadis and Fan (2001) which gives the necessary and sufficient conditions on a penalty function for the solution of the penalized least-suares problem to be thresholding, continuous and approximately unbiased when $|z|$ is large. Our purpose here is to show how to derive the penalties corresponding to the thresholding rules defined in the previous subsection, and to show that almost all of them satisfy these conditions. More precisely, we have

**Propostion 3.2.** *Let $\delta_\lambda : \mathbb{R} \to \mathbb{R}$ be a thresholding function that is increasing antisymmetric such that $0 \leq \delta_\lambda(x) \leq x$ for $x \geq 0$ and $\delta_\lambda(x) \to \infty$ as $x \to \infty$. Then there exist a continuous positive penalty function $p_\lambda$, with $p_\lambda(x) \leq p_\lambda(y)$ whenever $|x| \leq |y|$, such that $\delta_\lambda(z)$ is the unique solution of the minimization problem $\min_\theta (z - \theta)^2 + 2p_\lambda(|\theta|)$ for every $z$ at which $\delta_\lambda$ is continuous.*

*Proof.* Let $r_\lambda : \mathbb{R} \to \mathbb{R}$ defined by $r_\lambda(x) = \sup\{z | \delta_\lambda(z) \leq x\}$. The function $r_\lambda$ is well defined, since the set upon which the supremum is taken is non empty (recall that $\delta_\lambda(z) \to -\infty$ as $z \to -\infty$) and upper bounded (since $\delta_\lambda(z) \to \infty$ as $z \to \infty$. For $\theta \geq 0$, let

$$p_\lambda(\theta) = \int_0^\theta (r_\lambda(u) - u) du,$$

and suppose that $\delta_\lambda$ is continuous at $\theta$. Let

$$k(\theta) = (\theta - z)^2 + 2p_\lambda(\theta) - z^2 = 2 \int_0^\theta (r_\lambda(u) - z) du,$$

so that

$$k(\theta) - k(r_\lambda(z)) = 2 \int_{r_\lambda(z)}^\theta (r_\lambda(u) - z) du.$$

Using the assumptions, it is easy to show that for $\theta \neq r_\lambda(z)$, we have $k(\theta) > k(r_\lambda(z))$, so $\theta = \delta_\lambda(z)$ is the only solution to the minimization problem. The contracting property of the shrinkage function leads directly to the contracting property of the corresponding penalty. ☐

It is however interesting to recall here from the above proof the almost analytical expression for $p_\lambda$. Denote by $r_\lambda$ the generalized inverse function of $\delta_\lambda$ defined by $r_\lambda(x) = \sup\{z | \delta_\lambda(z) \leq x\}$. Then, for any $z > 0$, $p_\lambda$ is defined by

$$p_\lambda(z) = \int_0^z (r_\lambda(u) - u) du. \tag{3.14}$$

Note that all the thresholding function studied in the previous subsection satisfy the conditions of Proposition 3.2. Applying expression (3.14) one finds, in particular, the well known ridge regression $L_2$-penalty $p_\lambda(|\theta|) = \lambda |\theta|^2$ corresponding to the linear shrinkage function, the $L_1$-penalty $p_\lambda(|\theta|) = \lambda |\theta|$ corresponding to the soft thresholding rule and the hard thresholding penalty function $p_\lambda(|\theta|) = \lambda^2 - (|\theta| - \lambda)^2 I_{\{|\theta| < \lambda\}}(|\theta|)$ that results in the hard-thresholding rule (see Antoniadis (1997)). Figure 3 displays all penalty functions corresponding to the various shrinkage functions studied in this review.

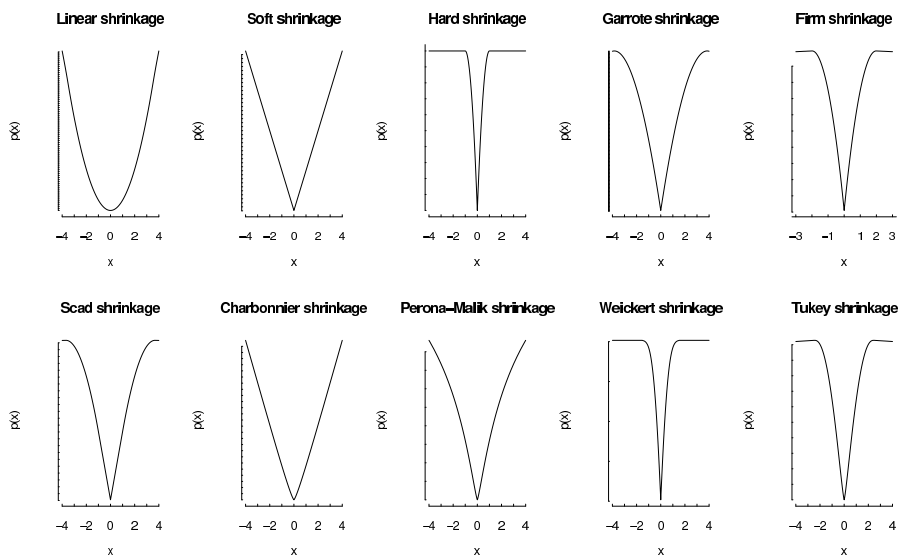

FIG 3. *Penalties corresponding to the shrinkage and thresholding functions with the same name.*

Among the penalties displayed in Figure 3, the quadratic penalty, while continuous is not singular at zero, and the resulting estimator is not thresholded. All other penalties are singular at zero, thus resulting in thresholding rules that enforce sparseness of the solution. The estimated wavelet coefficients obtained using the hard-thresholding penalty is not continuous at the threshold, so it may induce the oscillation of the reconstructed signal (lack of stability). In the soft-thresholding case, the resulting estimator of large coefficients is shifted by an amount of $\lambda$, which creates unnecessary bias when the coefficients are large. The same is true for Charbonnier and Perona-Malick penalties. All other penalties have similar behaviour. They are singular at zero (thus encouraging sparse solutions), continuous (thus stable) and do not create excessive bias when the wavelet coefficients are large. Most importantly, all the above thresholding penalties satisfy the conditions of Theorem 1 in Antoniadis and Fan (2001). The implication of this fact is that it leads to a systematic derivation of oracle inequalities and minimax properties for the resulting wavelet estimators via Theorem 2 of Antoniadis and Fan (2001). In particular, the optimal hard and soft universal threshold $\lambda = \sigma \sqrt{2 \log_2 n}$ given by Donoho and Johnstone (1994) leads to a sharp asymptotic risk upper bound and the resulting penalized estimators are adaptively minimax within a factor of logarithmic order over a wide range of Besov spaces.

### 3.3. Numerical examples

In this subsection, we illustrate the performance of the penalized least-squares methods introduced in the previous subsections by using some simulated data sets. For simulated data, we use the functions "heavisine", "blip", "corner" and "wave" as testing functions. The first three contain either jump points or cusps, while the fourth one is regular enough to see how the methods perform for regular functions. These, together with a typical Gaussian noisy data with a signal-to-noise ratio (SNR) of 3, are displayed in Figure 4.

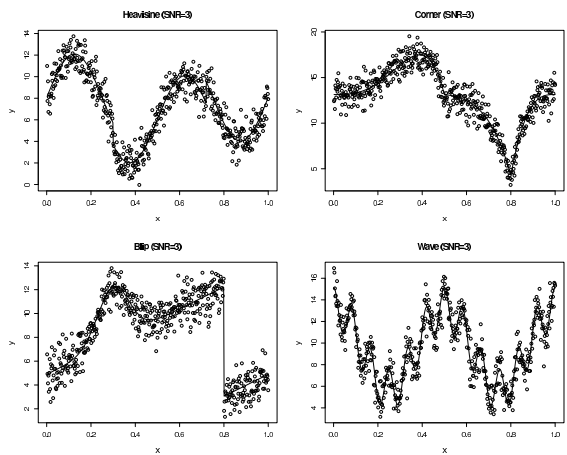

FIG 4. *The four signals used in the simulations (solid lines) together with a typical noisy version (points).*

For each signal, we have used two noise levels corresponding to signal-to-noise ratios 3 and 7 (low and high). For each simulation and each function, we have used an equispaced design of size 512 within the interval [0, 1] and a Gaussian noise was added to obtain the observed data. The experiment was repeated 100 times. Figure 5 dipslays a typical fit of the Heavisine function from corrupted data (SNR=3) and Table 1 reports the average mean-squared error over all Monte Carlo experiments for each method on each signal and signal-to-noise ratio combination.

As one can see most methods perform similarly when an universal threshold is used. Note also the bias of the soft-thresholding rule and the similarity between the firm and the scad shrinkage. From Table 1, we can see that the less classical nonlinear regularization methods are often superior to the hard-thresholding and the soft-thresholding method and always better than ridge regression.

### 3.4. Notes and remarks

In this whole section, we have tried to present a study for different denoising methods for signals observed on regularly spaced design and corrupted by ad-ditive Gaussian noise and have shown, that many of them are leading to the idea of shrinkage in a general sense. However, as stated already in the introduction, when the data does not meet these requirements (equally spaced and fixed sample points, periodic boundary conditions and i.i.d. normal errors) various modifications have been proposed in the recent literature and we would like to shortly mention few of them in this subsection.

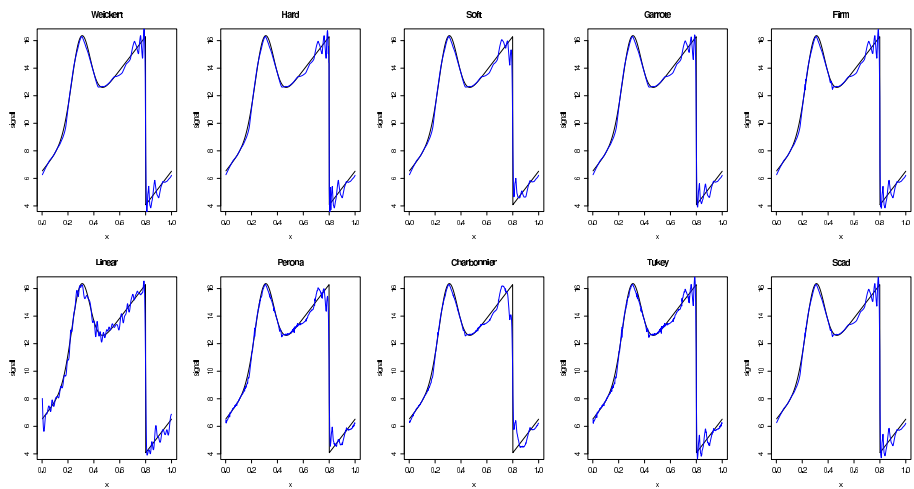

Fig 5. *A typical fit of the penalized least sqaures methods for the Blip function corrupted with an additive Gaussian noise at a signal-to-noise ratio of 3.*

TABLE 1
*Mean Squared Errors over 100 Simulations for each signal and SNR setting.*

| Method | Low SNR | | | | High SNR | | | |
|--------|-------|------|--------|------|-------|------|--------|------|
| | Heavi | Blip | Corner | Wave | Heavi | Blip | Corner | Wave |
| Weick | 80 | 95 | 63 | 212 | 46 | 64 | 27 | 147 |
| Hard | 65 | 76 | 60 | 152 | 48 | 57 | 30 | 147 |
| Soft | 113 | 181 | 67 | 472 | 46 | 108 | 26 | 249 |
| Garrote | 87 | 103 | 63 | 256 | 45 | 71 | 26 | 160 |
| Firm | 70 | 78 | 55 | 185 | 43 | 58 | 27 | 133 |
| Lin | 168 | 1849 | 69 | 4344 | 45 | 354 | 27 | 812 |
| Perona | 105 | 149 | 60 | 392 | 43 | 102 | 27 | 234 |
| Char | 136 | 354. | 656 | 9056 | 446 | 176 | 26 | 420 |
| Tukey | 84 | 90 | 57 | 247 | 42 | 65 | 28 | 129 |
| Scad | 82 | 90 | 58 | 241 | 43 | 65 | 27 | 140 |

An usual assumption underlying the use of wavelet shrinkage, that we have also adopted in the previous sections of this review, is that the regression function is assumed to be periodic, in order to use a wavelet transform with periodic boundary conditions. However, such an assumption is not always realistic. Oh and Lee (2005) propose an effective hybrid automatic method for correcting the boundary bias in wavelet shrinkage introduced by the inappropriateness of such a periodic assumption. Their idea is to combine wavelet shrinkage with local polynomial regression, where the latter regression technique is known to possess excellent boundary properties. Results from their numerical experiments strongly support this approach. The interested reader is referred to their paper for further details. Note that it is quite easy to implement the various thresholding strategies presented here within their hybrid approach.

For data that are not equispaced, there exist numerous wavelet methods in this setting. Among them let us cite Hall and Turlach (1997), Antoniadis et al. (1997), Cai and Brown (1998), Kovac and Silverman (2000), Antoniadis and Fan (2001), Maxim (2003), Chicken (2003), Kerkyacharian and Picard (2004) and Amato et al. (2006b). Compared to standard algorithms, the thresholding is notably more complicated because it has to incorporate the variations of the density of the design. In many of these constructions, the first step consists in determining a function $Y(x)$ of the form:

$$Y(x) = \sum_m w_m(x) Y_m,$$

where $w_m(x)$ is a sequence of functions suitably chosen. For instance, in Hall and Turlach (1997) the $w_m$s correspond to a polynomial which depends on the design. In Cai and Brown (1998) (and in Maxim (2003)), the $w_m$s corresponds to scale wavelets warped with the (known) design cumulative distribution function. In Antoniadis et al. (1997), the random design is transformed into equispaced data via a binning method and the weights $w_m$ are defined by scale wavelets. In a second step, the function $Y(x)$ is expanded on a standard wavelet basis and a hard thresholding algorithm is performed. In Antoniadis and Fan (2001) the $w_m$s are Sobolev interpolation weights. Kovac and Silverman (2000) apply first a linear transformation to the data to map it to a set of dyadic, equispaced points. Since the transformation induces a band-limited correlation on the resulting wavelet coefficients, they adopt a term-by-term thresholding method similar to VisuShrink, but taking into account the varying levels of noise. In Chicken (2003) linear transformations and block thresholding are applied to nondyadic, non-i.i.d., nonequispaced data with various band-limited covariance structures. His method makes use of Kovac and Silvermans fast algorithm for estimating the covariances of the transformed data. In all the techniques described above, the thresholds have similar forms and depend on the quantity $\sup_t \frac{1}{s(t)}$ where $s$ is the density of the design points, and which corresponds to an upper bound for the variance of the estimated wavelet coefficients. The algorithm of Amato et al. (2006b) is quite different and relies upon wavelet kernel reproducing Hilbert spaces. The regularization method suggested in their paper requires neither pre-processing of the data by interpolation or similar technique, nor the knowledge of the distribution $s$ of the design points. For this reason, the method works really well even in the case when the distribution of the design points deviates far from the uniform. When the estimation error is calculated at the design points only, the method achieves optimal convergence rates in Besov spaces no matter how irregular the design is. In order to obtain asymptotic optimality in the $L_2$ metric, an extra assumption on the design points should be imposed, namely, the density $s$ of the design points should be bounded away from zero. To end this paragraph we mention the algorithm developed in Kerkyacharian and Picard (2004). Their procedure stays very close to the equispaced Donoho and Johnstone's Visushrink procedure, and thus is very simple in its form (preliminary estimators are no longer needed) and in its implementation (the standard uniform threshold suffices). On the other side, the projection is done on an unusual non-orthonormal basis, called warped wavelet basis, so their analytic properties need to be studied to derive the performances of the estimator. Some theoretical results, including maxiset properties are established in their paper.

Concerning the noise, we would like to mention a recent work by Brown et al. (2006) who develop a nonparametric regression method that is simultaneously adaptive over a wide range of function classes for the regression function and robust over a large collection of error distributions, including those that are heavytailed, and may not even possess variances or means. Their approach is to first use local medians to turn the problem of nonparametric regression with unknown noise distribution into a standard Gaussian regression problem and then apply a wavelet block thresholding procedure to construct an estimator of the regression function. It is shown that the estimator simultaneously attains the optimal rate of convergence over a wide range of the Besov classes, without prior knowledge of the smoothness of the underlying functions or prior knowledge of the error distribution. The estimator also automatically adapts to the local smoothness of the underlying function, and attains the local adaptive minimax rate for estimating functions at a point. A key technical result in their development is a quantile coupling theorem which gives a tight bound for the quantile coupling between the sample medians and a normal variable.

We would like to add some comments concerning the choice of the penalty parameter $\lambda$ in finite sample situations. From a practical point of view its optimal choice is important. Given the basic framework of function estimation using wavelet thresholding and its relation with the regularization approach with penalties non smooth at 0, there are a variety of methods to choose the regularization parameter $\lambda$. Solo (2001) in his discussion of the paper by Antoniadis and Fan (2001) suggests a data based estimator of $\lambda$ similar in spirit to the SURE selection criterion used by Donoho and Johnstone (1994), and provides an appropriate simple SURE formula for the general penalties studied in this review. Another way to address the optimal choice of the regularization parameter is Generalized Cross Validation (GCV). Cross-validation has been widely used as an automatic procedure to choose the smoothing parameter in many statistical settings. The classical cross-validation method is performed by systematically expelling a data point from the construction of an estimate, predicting what the removed value would be and, then, comparing the prediction with the value of the expelled point. One way to proceed is to use the approach adopted by Jansen et al. (1997). It is clear that this is an area where further careful theoretical and practical work is needed.

### *3.5. Block thresholding for nonparametric regression*

In the previous subsections we have used separable penalties, which achieve minimax optimality through term-by-term thresholding of the empirical wavelet coefficients by realizing a degree of trade-off between variance and bias contribution to mean squared error. However this trade-off is not optimal. While separable rules have desirable minimax properties, they are necessarily not rate adaptive over Besov spaces under integrated mean squared error because they remove too many terms from the empirical wavelet expansion, with the result the estimator being too biased and having sub-optimal $L^2$-risk convergence rate (and also in $L^p$, $1 \leq p \leq \infty$). One way to increase estimation precision is by utilizing information about neighboring empirical wavelet coefficients. In other words, empirical wavelet coefficients could be thresholded in *blocks* (or groups) rather than individually. As a result, the amount of information available from the data for estimating the "average" empirical wavelet coefficient within a block, and making a decision about retaining or discarding it, would be an order of magnitude larger than the case of a term-by-term threshold rule. This would allow threshold decisions to be made more accurately and permit convergence rates to be improved. This is the spirit of block thresholding rules and blockwise additive penalty functions that have been and are currently theoretically explored by many researchers (see e.g. Hall et al. (1999), Cai and Brown (1999), Cai (2002), Cai and Silverman (2001), Chicken (2003) and Chicken and Cai (2005), Cai and Zhou (2005)). For completeness we present below some more details on such block-wise thresholding procedures.

### 3.5.1. A nonoverlapping block thresholding estimator

A nonoverlapping block thresholding estimator was proposed by Cai and Brown (1999) via the approach of ideal adaptation with the help of an oracle.

At each resolution level $j = j_0, \ldots, J - 1$, the empirical wavelet coefficients $\hat{d}_{jk}$ are grouped into nonoverlapping blocks of length $L$. In each case, the first few empirical wavelet coefficients might be re-used to fill the last block (which is called the *augmented* case) or the last few remaining empirical wavelet coefficients might not be used in the inference (which is called the *truncated* case), should $L$ not divide $2^j$ exactly.

Let $(jb)$ denote the set of indices of the coefficients in the $b$th block at level $j$, that is,

$$(jb) = \{(j,k) : (b-1)L + 1 \leq k \leq bL\},$$

and let $S^2_{(jb)}$ denote the $L^2$-energy of the noisy signal in the block $(jb)$. Within each block $(jb)$, estimate the wavelet coefficients $d_{jk}$ simultaneously via a James-Stein thresholding rule

$$\tilde{d}^{(jb)}_{jk} = \max \left( 0, \frac{S^2_{(jb)} - \lambda L \sigma^2}{S^2_{(jb)}} \right) \hat{d}_{jk}. \tag{3.15}$$

Then, an estimate of the unkown function $g$ is obtained by applying the IDWT to the vector consisting of both empirical scaling coefficients $\hat{c}_{j_0 k}$ ($k = 0, 1, \ldots, 2^{j_0} - 1$) and thresholded empirical wavelet coefficients $\tilde{d}^{(jb)}_{jk}$ ($j = j_0, \ldots, J - 1$; $k = 0, 1, \ldots, 2^j - 1$).

Cai and Brown (1999) suggested using $L = \log n$ and setting $\lambda = 4.50524$ which is the solution of the equation $\lambda - \log \lambda - 3 = 0$. This particular threshold was chosen so that the corresponding wavelet thresholding estimator is (near) optimal in function estimation problems. The resulting block thresholding estimator was called *BlockJS*.

**Remark 3.1.** *Hall et al. (1997) and Hall et al. (1998), Hall et al. (1999) considered wavelet block thresholding estimators by first obtaining a near unbiased estimate of the $L^2$-energy of the true coefficients whithin a block and then keeping or killing all the empirical wavelet coefficients within the block based on the magnitude of the estimate. Although it would be interesting to numerically compare their estimators, they require the selection of smoothing parameters – block length and threshold level – and it seems that no specific criterion is provided for choosing these parameters in finite sample situations.*

### 3.5.2. An overlapping block thresholding estimator

Cai and Silverman (2001) considered an overlapping block thresholding estimator by modifying the nonoverlapping block thresholding estimator of Cai (1999). The effect is that the treatment of empirical wavelet coefficients in the middle of each block depends on the data in the whole block.

At each resolution level $j = j_0, \ldots, J - 1$, one groups the empirical wavelet coefficients $\hat{d}_{jk}$ into nonoverlapping blocks $(jb)$ of length $L_0$. Extend each block by an amount $L_1 = \max(1, [L_0/2])$ in each direction to form overlapping larger blocks $(jB)$ of length $L = L_0 + 2L_1$.

Let $S^2_{(jB)}$ denote the $L^2$-energy of the noisy signal in the larger block $(jB)$. Within each block $(jb)$, estimate the wavelet coefficients simultaneously via the following James-Stein thresholding rule

$$\breve{d}^{(jb)}_{jk} = \max\left(0, \frac{S^2_{(jB)} - \lambda L \hat{\sigma}^2}{S^2_{(jB)}}\right) \hat{d}_{jk}. \tag{3.16}$$

Then, an estimate of the unkown function $g$ is obtained by applying the IDWT to the vector consisting of both empirical scaling coefficients $\hat{c}_{j_0 k}$ ($k = 0, 1, \ldots, 2^{j_0} - 1$) and thresholded empirical wavelet coefficients $\breve{d}^{(jb)}_{jk}$ ($j = j_0, \ldots, J - 1$; $k = 0, 1, \ldots, 2^j - 1$).

Cai and Silverman (2001) suggested using either $L_0 = [\log n/2]$ and taking $\lambda = 4.50524$ (which results in the *NeighBlock* estimator) or $L_0 = L_1 = 1$ (i.e., $L = 3$) and taking $\lambda = \frac{2}{3} \log n$ (which results in the *NeighCoeff* estimator). NeighBlock uses neighbouring coefficients outside the block of current interest in fixing the threshold, whilst NeighCoeff chooses a threshold for each coefficient by reference not only to that coefficient but also to its neighbours.

**Remark 3.2.** *The above thresholding rule (3.16) is different to the one given in (3.15) since the empirical wavelet coefficients $\hat{d}_{jk}$ are thresholded with reference to the coefficients in the larger block $(jB)$. One can envision $(jB)$ as a sliding* *window which moves $L_0$ positions each time and, for each window, only half of the coefficients in the center of the window are estimated.*

As noticed above, the block size and threshold level play important roles in the performance of a block thresholding estimator. The local block thresholding methods mentioned above all have fixed block size and threshold and the same thresholding rule is applied to all resolution levels regardless of the distribution of the wavelet coefficients. In a recent paper, Cai and Zhou (2005) propose a data-driven approach to empirically select both the block size and threshold at individual resolution levels. At each resolution level, their procedure, named SureBlock, chooses the block size and threshold empirically by minimizing Stein's Unbiased Risk Estimate (SURE). By empirically selecting both the block size and threshold and allowing them to vary from resolution level to resolution level, the SureBlock estimator has significant advantages over the more conventional wavelet thresholding estimators with fixed block sizes. For more details the reader is referred to the above paper.

We would like to end this subsection with some remarks on the use of block thresholding for density estimation by Chicken and Cai (2005). The reader interested in other wavelet based methods for density estimation from i.i.d. observations is referred to the review paper by Antoniadis (1997) where some linear and nonlinear wavelet estimators in univariate density estimation are discussed in detail. Chicken and Cai's block thresholding procedure first divides the empirical coefficients at each resolution level into nonoverlapping blocks and then simultaneously keeps or kills all the coefficients within a block, based on the magnitude of the sum of the squared empirical coefficients within that block. Motivated by the analysis of block thresholding rules for nonparametric regression, the block size is chosen to be $\log n$. It is shown that the block thresholding estimator adaptively achieves not only the optimal global rate over Besov spaces, but simultaneously attains the adaptive local convergence rate as well. These results are obtained in part through the determination of the optimal block length.

## 4. Some applications

In this Section we present two recent applications of wavelets in statistics. Each application begins with a description of the problem under study and points to specific properties and techniques which were used to determine a solution. Of course many other excellent applied works on wavelets have been presented in the literature that we find it impossible to list them all in this review. We will mainly concentrate on some important uses of wavelet methods in statistics, developed recently by the author and its collaborators.

### *4.1. Wavelet thresholding in partial linear models*

This subsection is concerned with a semiparametric partially linear regression model with unknown regression coefficients, an unknown nonparametric func-tion for the non-linear component, and unobservable Gaussian distributed random errors. We present a wavelet thresholding based estimation procedure to estimate the components of the partial linear model by establishing the connection made between an $L_1$-penalty based wavelet estimator of the nonparametric component and Huber 's M-estimation of a standard linear model with outliers.

Assume that responses $y_1, \ldots, y_n$ are observed at deterministic equidistant points $t_i = \frac{i}{n}$ of an univariate variable such as time and for fixed values $\mathbf{X}_i$, $i = 1, \ldots, n$, of some $p$-dimensional explanatory variable and that the relation between the response and predictor values is modeled by a Partially Linear Model (PLM):

$$y_i = \mathbf{X}_i^T \boldsymbol{\beta}_0 + f(t_i) + u_i \qquad i = 1 \ldots n, \tag{4.1}$$

where $\boldsymbol{\beta}_0$ is an unknown $p$-dimensional real parameter vector and $f$ is an unknown real-valued function; the $u_i$'s are i.i.d. normal errors with mean 0 and variance $\sigma^2$ and superscript "T" denotes the transpose of a vector or matrix. We will assume hereafter that the sample size is $n = 2^J$ for some positive integer $J$. Given the observed data $(y_i, \mathbf{X}_i)_{i=1 \ldots n}$, the aim is to estimate from the data the vector $\boldsymbol{\beta}$ and the function $f$.

Partially linear models are more flexible than the standard linear model because they combine both parametric and nonparametric components when it is believed that the response variable $Y$ depends on variable $\mathbf{X}$ in a linear way but is nonlinearly related to other independent variable $t$. As it is well known, model 4.1 is often used when the researcher knows more about the dependence among $y$ and $\mathbf{X}$ than about the relationship between $y$ and the predictor $t$, which establishes an unevenness in prior knowledge. Several methods have been proposed in the literature to consider partially linear models. One approach to estimation of the nonparametric component in these models is based on smoothing splines regression techniques and has been employed in particular by Green and Yandell (1985), Engle et al. (1986) and Rice (1986) among others. Kernel regression (see e.g. Speckman (1988)) and local polynomial fitting techniques (see e.g. Hamilton and Truong (1997)) have also been used to study partially linear models. An important assumption by all these methods for the unknown nonparametric component $f$ is its high smoothness. But in reality, such a strong assumption may not be satisfied. To deal with cases of a less-smooth nonparametric component, a wavelet based estimation procedure, developed recently by Gannaz (2006) (a PhD student of the author), is developed in what follows, and as such that it can handle nonparametric estimation for curves lying in Besov spaces instead of the more classical Sobolev spaces. To capture key characteristics of variations in $f$ and to exploit its sparse wavelet coefficients representation, we assume that $f$ belongs to $\mathcal{B}_{\pi,r}^s([0;1])$ with $s + 1/\pi - 1/2 > 0$. The last condition ensures in particular that evaluation of $f$ at a given point makes sense.

### 4.1.1. Estimation procedure

In matrix notation, the PLM model specified by (4.1) can be written as

$$\mathbf{Y} = X\boldsymbol{\beta}_0 + \mathbf{F} + \mathbf{U}, \tag{4.2}$$

where $\mathbf{Y} = (Y_1, \ldots, Y_n)^T$, $X^T = (\mathbf{X}_1, \ldots, \mathbf{X}_n)$ is the $p \times n$ design matrix, and $\mathbf{F} = (f(t_1), \ldots f(t_n))^T$. The noise vector $\mathbf{U} = (u_1, \ldots, u_n)^T$ is a Gaussian vector with mean 0 and variance matrix $\sigma^2 I_n$.

Let now $\mathbf{Z} = W\mathbf{Y}$, $A = WX$, $\boldsymbol{\theta} = W\mathbf{F}$ and $\boldsymbol{\epsilon} = W\mathbf{U}$, where $W$ is the discrete wavelet transform operator. Pre-multiplying (4.1) by $W$, we obtain the transformed model

$$\mathbf{Z} = A\boldsymbol{\beta}_0 + \boldsymbol{\theta}_0 + \boldsymbol{\epsilon}. \tag{4.3}$$

The orthogonality of the DWT matrix $W$ ensures that the transformed noise vector $\boldsymbol{\epsilon}$ is still distributed as a Gaussian white noise with variance $\sigma^2 I_n$. Hence, the representation of the model in the wavelet domain not only allows to retain the partly linear structure of the model but also to exploit in an efficient way the sparsity of the wavelet coefficients in the representation of the nonparametric component.

With the basic understanding of wavelet based penalized least squares procedures covered in depth in Section 3, we propose estimating the parameters $\boldsymbol{\beta}$ and $\boldsymbol{\theta}$ in model (4.3) by penalized least squares. To be specific, our wavelet based estimators are defined as follows:

$$(\hat{\boldsymbol{\beta}}_n, \hat{\boldsymbol{\theta}}_n) = \underset{(\boldsymbol{\beta}, \boldsymbol{\theta})}{argmin} \left\{ J_n(\boldsymbol{\beta}, \boldsymbol{\theta}) = \sum_{i=1}^n \frac{1}{2}(z_i - A_i^T \boldsymbol{\beta} - \boldsymbol{\theta}_i)^2 + \lambda \sum_{i=i_0}^n |\boldsymbol{\theta}_i| \right\}, \tag{4.4}$$

for a given penalty parameter $\lambda$, where $i_0 = 2^{j_0} + 1$. The penalty term in the above expression penalizes only the empirical wavelet coefficients of the nonparametric part of the model and not its scaling coefficients. Remember that the $L^1$-penalty is associated the soft thresholding rule.

The regularization method proposed above is closely related to the method proposed recently by Chang and Qu (2004), who essentially concentrate on the backfitting algorithms involved in the optimization, without any theoretical study of the resulting estimates. The method also relates to the recent one developed by Fadili and Bullmore (2005) where a variety of penalties is discussed. Note, however, that their asymptotic study is limited to quadratic penalties which amounts essentially in assuming that the underlying function $f$ belongs to some Sobolev space.

Let us now have a closer look at the minimization of the criterion $J_n$ stated in (4.4). For a fixed value of $\boldsymbol{\beta}$, the criterion $J_n(\boldsymbol{\beta}, \cdot)$ is minimum at

$$\tilde{\theta}_i(\boldsymbol{\beta}) = \begin{cases} z_i - A_i^T \boldsymbol{\beta} & \text{if } i < i_0, \\ \text{sign}(z_i - A_i^T \boldsymbol{\beta}) \left( |z_i - A_i^T \boldsymbol{\beta}| - \lambda \right)_+ & \text{if } i \geq i_0. \end{cases} \tag{4.5}$$

Therefore, finding $\hat{\boldsymbol{\beta}}_n$, a solution to problem (4.4), amounts in finding $\hat{\boldsymbol{\beta}}_n$ minimizing the criterion $J_n(\boldsymbol{\beta}, \tilde{\boldsymbol{\theta}}(\boldsymbol{\beta}))$. However, note that

$$J_n(\boldsymbol{\beta}, \tilde{\boldsymbol{\theta}}(\boldsymbol{\beta})) = \sum_{i=i_0}^{n} \rho_\lambda(z_i - A_i^T \boldsymbol{\beta}) \quad (4.6)$$

where $\rho_\lambda$ is Huber's cost functional with threshold $\lambda$, defined by:

$$\rho_\lambda(u) = \begin{cases} u^2/2 & \text{if } |u| \le \lambda, \\ \lambda|u| - \lambda^2/2 & \text{if } |u| > \lambda. \end{cases} \quad (4.7)$$

The above facts can be derived as follows. Let $i \ge i_0$. Minimizing expression (4.4) with respect to $\theta_i$ is equivalent in minimizing $j(\theta_i) := \frac{1}{2}(z_i - A_i^T \boldsymbol{\beta} - \theta_i)^2 + \lambda|\theta_i|$. The first order conditions for this are : $j'(\theta_i) = \theta_i - (z_i - A_i^T \boldsymbol{\beta}) + \text{sign}(\theta_i)\lambda = 0$ where $j'$ denotes the derivative of $j$. Now,

- if $\theta_i \ge 0$, then $j'(\theta_i) = 0$ if and only if $\theta_i = z_i - A_i^T \boldsymbol{\beta} - \lambda$. Hence, if $z_i - A_i^T \boldsymbol{\beta} \le \lambda$, $\theta_i = 0$ and otherwise $\theta_i = z_i - A_i^T \boldsymbol{\beta} - \lambda$.
- if $\theta_i \le 0$, $j'(\theta_i)$ is zero if and only if $\theta_i = z_i - A_i^T \boldsymbol{\beta} + \lambda$; therefore, if $z_i - A_i^T \boldsymbol{\beta} \ge -\lambda$, $\theta_i = 0$ and otherwise $\theta_i = z_i - A_i^T \boldsymbol{\beta} + \lambda$.

This proves that for a fixed value of $\boldsymbol{\beta}$, the criterion (4.4) is minimal for $\tilde{\boldsymbol{\theta}}(\boldsymbol{\beta})$ given by expression (4.5). If we now replace $\boldsymbol{\theta}$ in the objective function $J_n$ we obtain $J_n(\boldsymbol{\beta}, \tilde{\boldsymbol{\theta}}(\boldsymbol{\beta})) = \frac{1}{2} \sum_{i=i_0}^{n} \left( (z_i - A_i^T \boldsymbol{\beta} - \tilde{\theta}_i)^2 + \lambda|\tilde{\theta}_i| \right)$ since $\tilde{\theta}_i = z_i - A_i^T \boldsymbol{\beta}$ for $i < i_0$. Now denoting by $I$ the set $I := \{j = i_0 \dots n, \ |z_j - A_j \boldsymbol{\beta}| < \lambda\}$, we find that $J_n(\boldsymbol{\beta}, \tilde{\boldsymbol{\theta}}(\boldsymbol{\beta})) = \frac{1}{2} \sum_I (z_i - A_i^T \boldsymbol{\beta})^2 + \frac{1}{2} \sum_{I^C} \lambda^2 + \lambda \sum_{I^C} \left( |z_i - A_i^T \boldsymbol{\beta}| - \lambda \right)$ by replacing $\tilde{\theta}_i$ with (4.5), which is exactly Huber's functional.

This result allows the computation of the estimators $\hat{\boldsymbol{\beta}}_n$ et $\hat{\boldsymbol{\theta}}_n$ in a non-iterative fashion. The resulting form of the estimators allows us to study their asymptotic properties (see Gannaz (2006)). Another benefit is that one can design estimation algorithms much faster than those based on backfitting.

The estimation procedure may be summarized as follows. Using the observed data $(\mathbf{Y}, X)$ :

1. Apply the DWT of order $J = \log_2(n)$ on $X$ and $\mathbf{Y}$ to get their corresponding representation $A$ and $\mathbf{Z}$ in the wavelet domain.
2. The parameter $\boldsymbol{\beta}_0$ is then Huber's robust estimator which is obtained without taking care of the nonparametric component in the PLM model:

$$\hat{\boldsymbol{\beta}}_n = \underset{\boldsymbol{\beta}}{argmin} \sum_{i=1}^{n} \rho_\lambda(z_i - A_i^T \boldsymbol{\beta}).$$

In other words this amounts in considering the linear model $z_i = A_i^T \boldsymbol{\beta}_0 + e_i$ with noise $e_i = \theta_{0,i} + \epsilon_i$.

3. The vector $\boldsymbol{\theta}$ of wavelet coefficients of the function $f$ is estimated by soft thresholding of $Z - A\hat{\boldsymbol{\beta}}_n$:

$$\hat{\theta}_{i,n} = \begin{cases} z_i - A_i^T \hat{\boldsymbol{\beta}}_n & \text{if } i < i_0, \\ \text{sign}(z_i - A_i^T \hat{\boldsymbol{\beta}}_n) \left( |z_i - A_i^T \hat{\boldsymbol{\beta}}_n| - \lambda \right)_+ & \text{if } i \geq i_0. \end{cases}$$

The estimation of $f$ is then obtained by applying the inverse discrete wavelet transform. Note that this last step corresponds to a standard soft-thresholding nonparametric estimation of $f$ in the model:

$$Y_i - \mathbf{X}_i^T \hat{\boldsymbol{\beta}}_n = f(t_i) + v_i, \quad i = 1 \ldots n,$$

where $v_i = \mathbf{X}_i^T (\boldsymbol{\beta}_0 - \hat{\boldsymbol{\beta}}_n) + u_i$.

**Remark 4.1.** *The wavelet soft-thresholding procedure proposed above is derived by exploiting the connection between an $L_1$-based penalization of the wavelet coefficients of $f$ and Huber's M-estimators in a linear model. Other penalties, leading to different thresholding procedures can also be seeing as M-estimation procedures. For example, if $\delta_\lambda$ denotes the resulting thresholding function, we can show in a similar way that the estimators verify*

$$\begin{aligned} \hat{\boldsymbol{\beta}}_n &= \underset{\boldsymbol{\beta}}{argmin} \sum_{i=i_0}^n \rho_\lambda(z_i - A_i^T \boldsymbol{\beta}), \\ \hat{\theta}_{i,n} &= \begin{cases} z_i - A_i^T \boldsymbol{\beta} & \text{if } i < i_0, \\ \delta_\lambda(z_i - A_i^T \boldsymbol{\beta}) & \text{if } i \geq i_0, \end{cases} \quad i = 1 \ldots n, \end{aligned}$$

*with $\rho_\lambda$ being the primitive of $u \mapsto u - \delta_\lambda(u)$. From the above, one sees that hard thresholding corresponds to mean truncation, while SCAD thresholding is associated to Hampel's M-estimation. However, in this subsection, we only concentrate on the properties of estimators obtained by soft thresholding.*

Existing results for semi-parametric partial linear models establish parametric rates of convergence for the linear part and minimax rates for the nonparametric part, showing in particular that the the existence of a linear component does not changes the rates of convergence of the nonparametric component. Within the framework adopted in this paper, the rates of convergence are similar, but an extra logarithmic term appears in the rates of the parametric part, mainly due to the fact that the smoothness assumptions on the nonparametrric part are weaker. Indeed, under appropriate assumptions, one has:

**Propostion 4.1.** *Let $\hat{\boldsymbol{\beta}}_n$ and $\hat{\boldsymbol{\theta}}_n$ be the estimators defined above. Consider that the penalty parameter $\lambda$ is the universal threshold: $\lambda = \sigma\sqrt{2\log(n)}$. Then it holds*

$$\hat{\boldsymbol{\beta}}_n - \boldsymbol{\beta}_0 = O_P\left( \sqrt{\frac{\log(n)}{n}} \right).$$

*If in addition we assume that the scaling function $\phi$ and the mother wavelet $\psi$ belong to $\mathcal{C}^R$ and that $\psi$ has $N$ vanishing moments, then, for $f$ belonging to the Besov space $\mathcal{B}_{\pi,r}^s$ with $0 < s < \min(R, N)$, one has*

$$\|\hat{f}_n - f\|_2 = O_P\left(\left(\frac{\log(n)}{n}\right)^{\frac{1}{1+2s}}\right),$$

*where $\|\hat{f}_n - f\|_2^2 = \int_0^1 (\hat{f}_n - f)^2$.*

A proof of this proposition together with the relevant assumptions may be found in Gannaz (2006). The presence of a logarithmic loss lies on the choice of the threshold $\lambda$: taking $\lambda$ which tends to 0, as suggested by Fadili and Bullmore (2005), would lead to a minimax rate in the estimation of $\boldsymbol{\beta}$. The drawback is that the quality of the estimation for the nonparametric part of the PLM would not be anymore quasi-minimax. This phenomenon was put in evidence by Rice (1986): a compromise must be done between the optimality of the linear part estimation with an oversmoothing of the functional estimation and a loss in the linear regression parameter convergence rate but a correct smoothing of the functional part.

### 4.1.2. Estimation of the variance

The estimation procedure relies upon knowledge of the variance $\sigma^2$ of the noise, appearing in the expression of the threshold $\lambda$. In practice, this variance is unknown and needs to be estimated. In wavelet approaches for standard nonparametric regression, a popular and well behaved estimator for the unknown standard deviation of the noise is the median absolute deviation (MAD) of the finest detail coefficients of the response divided by 0.6745 (see Donoho et al. (1995)). The use of the MAD makes sense provided that the wavelet representation of the signal to be denoised is sparse. However, such an estimation procedure cannot be applied without some pretreatment of the data in a partially linear model because the wavelet representation of the linear part of a PLM may be not sparse.

A QR decomposition on the regression matrix of the PLM allow to eliminate this bias. Since often the function wavelet coefficients at weak resolutions are not sparse, we only consider the wavelet representation at a level $J = \log_2(n)$. Let $A_J$ be the wavelet representation of the design matrix $X$ at level $J$. The QR decomposition ensures that there exist an orthogonal matrix $Q$ and an upper triangular matrix $R$ such that

$$A_J = Q\begin{pmatrix} R \\ 0 \end{pmatrix}.$$

If $\mathbf{Z}_J$, $\boldsymbol{\theta}_{0,J}$ and $\boldsymbol{\epsilon}_J$ denote respectively the vector of the wavelets coefficients at resolution $J$ of $Y$, $F_0$ and $U$, model (4.3) gives

$$Q^T \mathbf{z}_J = \begin{pmatrix} R \\ 0 \end{pmatrix}\boldsymbol{\beta}_0 + Q^T \boldsymbol{\theta}_{0,J} + Q^T \boldsymbol{\epsilon}_J.$$

It is easy to see that applying the MAD estimation on the last components of $Q^T \mathbf{z}_J$ rather than on $\mathbf{z}_J$ leads to a satisfactory estimation of $\sigma$. Indeed thanks to the QR decomposition the linear part does not appear anymore in the estimation and thus the framework is similar to the one used in nonparametric regression. Following Donoho and Johnstone (1998), the sparsity of the functional part representation ensures good properties of the resulting estimator.

The interested reader will find in Gannaz (2006) two particular optimization algorithms that may be used for estimating in a computationally efficient way the linear part of the model.

## *4.2. Dimension reduction in functional regression*

In functional regression problems, one has a response variable $Y$ to be predicted by a set of variables $X_1, \ldots, X_p$ that are discretizations of a same curve $X$ at points $t_1, \ldots, t_p$, that is $X_j = X(t_j)$, $j = 1, \ldots, p$ where the discretization time points $t_j$ lie in $[0, 1]$ without loss of generality. A typical set-up includes near infra-red spectroscopy ($Y$ represents the proportion of a chemical constituent and $\mathbf{x}$ is the spectrum of a sample, discretized at a sequence of wavelengths).

In practice, before applying any nonparametric regression technique to model real data, and in order to avoid the curse of dimensionality, a dimension reduction or model selection technique is crucial for appropriate smoothing. A possible approach is to find a functional basis, decompose the covariate curve $X(t)$ accordingly and work with the coefficients in a spirit similar to that adopted by Martens and Naes (1989), chapter 3, Alsberg (1993) and Denham and Brown (1993). The aim is to explain or predict the response through an expansion of the explanatory process in a relatively low dimensional basis in the space spanned by the measurements of $X$, thus revealing relationships that may not be otherwise apparent. Dimension reduction without loss of information is a dominant theme in such cases: one tries to reduce the dimension of $X$ without losing information on $Y|X$, and without requiring a model for $Y|X$. Borrowing terminology from classical statistics, Cook (2000) calls this a sufficient dimension reduction which leads to the pursuit of sufficient summaries containing all of the information on $Y|X$ that is available from the sample.

In this subsection we describe a wavelet based regression approach for regression problems with high dimensional $X$ variables, developed recently by the author and its co-authors (see Amato et al. (2006a)). The methodology relies upon the above notion of sufficient dimension reduction and is based on developments in a recent paper by Xia et al. (2002) where an adaptive approach for effective dimension reduction called the (conditional) minimum average variance estimation (MAVE) method, is proposed within quite a general setting.

### *4.2.1. Wavelet based MAVE*

We describe below the application of the MAVE method via a wavelet decomposition of the explanatory covariate process. We suppose that each realization $X(t)$ of the covariate process will be modelled as $X(t) = f(t) + s(t)$, $t \in [0, 1]$, where $f(t)$ represents the mean at time $t$ and $s(t)$ is the observed residual variation, which will be regarded as a realization of a second order weakly stationary process. Since a large number of signal compression algorithms are based on second order statistical information we will concentrate on covariance modelling, and the mean function will be removed by filtering or simple averaging. Thus, we assume hereafter that the covariate process has been centered, so that $E(X(t)) = 0$ for all $t$.

Consider then the following model

$$Y = g\left(\langle \beta_1, X \rangle, \ldots, \langle \beta_K, X \rangle\right) + \varepsilon \tag{4.8}$$

where $\varepsilon$ is a scalar random variable independent of $X(t)$, $\{\beta_s(t), s = 1, \ldots, K\}$ are $K$ orthonormal functions belonging to $L^2([0,1])$, and $g$ is a smooth link function of $\mathbb{R}^K$ into $\mathbb{R}$. For $D = K$, we obtain a standard regression model with all explanatory variables $\langle \beta_s, X \rangle$, $s = 1, \ldots, K$ entering independently. Provided that $D < K$, the regression function depends on $X$ only through $D$ linear functionals of the explanatory process $X$. Hence, to explain the dependent variable $Y$, the space of $K$ explanatory variables can be reduced to a space with a smaller dimension $D$. The dimension reduction methods aim to find the dimension $D$ of the reduction space and a basis defining this space.

Given a multiresolution analysis of $L^2([0,1])$ and a primary level $j_0$, as seen in Section 2, both $X(t)$ and $\beta_s(t)$ can be decomposed as

$$
\begin{aligned}
X(t) &= \sum_{k=0}^{2^{j_0}-1} \xi_{j_0,k} \phi_{j_0,k} + \sum_{j \geq j_0} \sum_{k=0}^{2^j - 1} \eta_{j,k} \psi_{j,k} \\
\beta_s(t) &= \sum_{k=0}^{2^{j_0}-1} c_{j_0,k}^s \phi_{j_0,k} + \sum_{j \geq j_0} \sum_{k=0}^{2^j - 1} d_{j,k}^s \psi_{j,k}
\end{aligned}
$$

with

$$
\begin{aligned}
\xi_{j_0,k} &= \langle X, \phi_{j_0,k} \rangle &\text{and} &&\eta_{j,k} &= \langle X, \psi_{j,k} \rangle \\
c_{j_0,k}^s &= \langle \beta_s, \phi_{j_0,k} \rangle &\text{and} &&d_{j,k}^s &= \langle \beta_s, \psi_{j,k} \rangle,
\end{aligned}
$$

$s = 1, \ldots, K$ and $\{\xi_{j_0,k}, \eta_{j,k}\}_{j,k}$ sequences of random variables. By the isometry between $L^2([0,1])$ and $\ell^2(\mathbb{R})$, model (4.8) can be also written as

$$Y = g\left(\langle \boldsymbol{\beta}_1, \boldsymbol{\gamma} \rangle, \ldots, \langle \boldsymbol{\beta}_K, \boldsymbol{\gamma} \rangle\right) + \varepsilon. \tag{4.9}$$

We have indicated by $\boldsymbol{\beta}_s$ the $\ell^2$-sequence formed by the wavelet and scaling coefficients of $\beta_s(t)$, $s = 1, \ldots, K$; and by $\boldsymbol{\gamma}$ the $\ell^2$-sequence formed by the wavelet and scaling coefficients of $X(t)$.

Usually, the sample paths of the process $X$ are discretized. If we observe $p = 2^J$ values of $X(t)$, $(X_1, \ldots, X_p) = (X(t_1), \ldots, X(t_p))$, then, given the previous notation, $X(t)$ can be approximated by its 'empirical' projection onto the approximation space $V_J$ defined as

$$X^J(t) = \sum_{k=0}^{2^{j_0}-1} \tilde{\xi}_{j_0,k} \phi_{j_0,k} + \sum_{j=j_0}^{J-1} \sum_{k=0}^{2^j-1} \tilde{\eta}_{j,k} \psi_{j,k}$$

where $\tilde{\xi}_{j_0,k}$ and $\tilde{\eta}_{j,k}$ are the empirical scaling and wavelet coefficients. We will collect them into a vector $\tilde{\boldsymbol{\gamma}}^J \in \mathbb{R}^p$. Let $\beta_s^J$ be the projection of $\beta_s$ onto $V_J$ and let us denote by $\boldsymbol{\beta}_s^J \in \mathbb{R}^p$ the vector collecting its scaling and wavelet coefficients, $s = 1, \ldots, K$. The original model (4.9) is then replaced by its discrete counterpart

$$Y = g\left(\langle \boldsymbol{\beta}_1^J, \tilde{\boldsymbol{\gamma}}^J \rangle, \ldots, \langle \boldsymbol{\beta}_K^J, \tilde{\boldsymbol{\gamma}}^J \rangle\right) + \varepsilon. \tag{4.10}$$

As much as $K$ and $J$ are large enough and thanks to the isometries between $L_2$ and $\ell_2$ and the compression properties of the wavelet transform, the original functional regression model (4.8) may be replaced by the above model (4.10) which is the candidate for further dimension reduction by MAVE. A compact way to write down model (4.10) is

$$Y = g\left(B^T \tilde{\boldsymbol{\gamma}}^J\right) + \varepsilon, \tag{4.11}$$

$B \in \mathbb{R}^{p \times K}$, $p = 2^J$. The method is then applied to Eq. (4.11). For the sake of completeness, we briefly describe hereafter the MAVE method, as it is applied on data obeying model (4.11).

Let $d$ represent now the working dimension, $1 \le d \le K$. For an assumed number $d$ of directions in model (4.10) and known directions $B_0$, one would typically minimize

$$\min \mathbb{E}\{Y - E(Y|B_0^T \tilde{\boldsymbol{\gamma}}^J)\}^2,$$

to obtain a nonparametric estimate of the regression function $\mathbb{E}(Y|B_0^T \tilde{\boldsymbol{\gamma}}^J)$. The MAVE method is based on the local linear regression, which hinges at a point $\tilde{\boldsymbol{\gamma}}_0^J$ on linear approximation

$$\mathbb{E}(Y|B_0^T \tilde{\boldsymbol{\gamma}}^J) \approx a + b^T B_0^T (\tilde{\boldsymbol{\gamma}}^J - \tilde{\boldsymbol{\gamma}}_0^J). \tag{4.12}$$

Now, if directions $B_0$ are not known, we have to search their approximation $B$. Xia et al. (2002) propose to plug-in unknown directions $B$ in the local linear approximation of the regression function and to optimize simultaneously with respect to $B$ and local parameters $a$ and $b$ of local linear smoothing. Hence, given a sample $(\tilde{\boldsymbol{\gamma}}_i^J, Y_i)_{i=1}^n$ from $(\tilde{\boldsymbol{\gamma}}^J, Y)$, they perform local linear regression at every $\tilde{\boldsymbol{\gamma}}_0^J = \tilde{\boldsymbol{\gamma}}_i^J$, $i = 1, \ldots, n$, and end up minimizing

$$\min_{\substack{B : B^T B = I_K \\ a_l, b_l, l = 1, \ldots, n}} \sum_{l=1}^n \sum_{i=1}^n \left\{ Y_i - \left[a_l + b_l^t B^T \left(\tilde{\boldsymbol{\gamma}}_i^J - \tilde{\boldsymbol{\gamma}}_l^J\right)\right]\right\}^2 w_{il} \tag{4.13}$$

where $I_K$ represents the $K \times K$ identity matrix and $w_{il}$ are weights describing the local character of the linear approximation (4.12) (i.e., $w_{il}$ should depend on the distance of points $\tilde{\boldsymbol{\gamma}}_i^J$ and $\tilde{\boldsymbol{\gamma}}_l^J$).

Xia et al. (2002) call the resulting estimator of $B$, MAVE and show that the simultaneous minimisation with respect to local linear approximation given by $a_j$, $b_j$ and to directions $B$ results in a convergence rate superior to any other dimension-reduction method. Initially, a natural choice of weights is given by a multidimensional kernel function $K_h$. At a given $\tilde{\gamma}_0^J$,

$$w_{i0} = K_h(\tilde{\gamma}_i^J - \tilde{\gamma}_0^J) / \sum_{i=1}^n K_h(\tilde{\gamma}_i^J - \tilde{\gamma}_0^J),$$

for $i = 1, \ldots, n$ and a kernel function $K_h(\cdot)$, where $h$ refers to a bandwidth. Additionally, when we already have an initial estimate of the dimension reduction space given by $\hat{B}$, it is possible to iterate and search an improved estimate of the reduction space. Xia et al. (2002) do so by using the initial estimator $\hat{B}$ to measure distances between points $\tilde{\gamma}_i^J$ and $\tilde{\gamma}_0^J$ in the reduced space. More precisely, they propose to choose in the iterative step weights

$$w_{i0} = K_h(\hat{B}^T(\tilde{\gamma}_i^J - \tilde{\gamma}_0^J)) / \sum_{i=1}^n K_h(\hat{B}^T(\tilde{\gamma}_i^J - \tilde{\gamma}_0^J)).$$

Repeating such iteration steps until convergence results in a refined MAVE (rMAVE) estimator. As one sees from the above equations, the initial estimate $\hat{B}$ depends on local linear smoothing performed with weights computed via a multidimensional kernel on $\mathbb{R}^p$. Since, by Theorem 1 in Xia et al. (2002) the optimal kernel bandwidth $h$ must be such that $h = \mathcal{O}(\log n / n^{1/p})$, in order to avoid the curse of dimensionality and to stabilize the computations it is therefore advisable in practice to reduce the initial resolution $\log_2 p$ to some resolution $J < \log_2 p$, and in such a way that the approximation of $\tilde{\gamma}$ by its projection $\tilde{\gamma}^J$ does not affect the asymptotics. When assuming that the process $X$ is $\alpha$-regular such a condition holds if $2^\alpha J = \mathcal{O}(n^{1/p})$ (see Amato et al. (2006a)). From now on, whenever we refer to MAVE, we mean its refined version rMAVE with such a choice of smoothing parameters. One may show that under appropriate assumptions on $X$ and $g$ and with such a choice of smoothing parameters, one gets optimal asymptotic rates. The interested reader is referred to the paper of Amato et al. (2006a) for more details.

The described methods are capable of estimating the dimension reduction space provided we can specify its dimension. To determine the dimension $d$, Xia et al. (2002) extend the cross-validation approach of Yao and Tong (1994). The cross-validation criterion is defined as

$$CV(d) = \sum_{j=1}^n \left[ Y_j - \sum_{i=1, i \neq j}^n \frac{Y_i K_h(\hat{B}^T(\tilde{\gamma}_i^J - \tilde{\gamma}_j^J))}{\sum_{i=1, i \neq j}^n Y_i K_h(\hat{B}^T(\tilde{\gamma}_i^J - \tilde{\gamma}_j^J))} \right],$$

for $d > 0$ and for the special case of independent $Y$ and $X$ as

$$CV(0) = \frac{1}{n} \sum_{i=1}^n (Y_i - \bar{Y})^2.$$

Consequently, the dimension is then determined as

$$\hat{d} = \underset{0 \le d \le K}{argmin}\, CV(d),$$

where $K$ represents the initial number of basis functions in model (4.8).

### 4.2.2. Numerical experiment

To illustrate the performance of the dimensional reduction regression method proposed in this subsection we report here some of the results of an extensive Monte Carlo simulation study realized in Amato et al. (2006a) for a particular model. More precisely, let $H = L^2([0,1])$, $\mathbf{X} = (X(t))_{t \in [0,1]}$ be a standard Brownian motion and $\epsilon$ a mean zero Gaussian distribution with variance $\sigma^2$ independent of $\mathbf{X}$. All curves $\beta_i(t)$ and $X(t)$ are discretized on the same grid generated on an equispaced grid of $p = 128$ equispaced points $t \in [0,1]$. The observations $Y$ are generated from i.i.d. observations of $X$ and $\epsilon$ according to the following model:

**Model 1.**

$$Y = \exp\left(\langle \beta_1, X \rangle\right) + \exp\left(|\langle \beta_2, X \rangle|\right) + \epsilon,$$

$\beta_1(t) = \sin(3\pi t/2)$, $\beta_2(t) = \sin(5\pi t/2)$, $\sigma = 0.1$

The motivation for this example is that the functions $\beta_1$ and $\beta_2$ belong to the eigen-subspace of the covariance operator of $X$ and therefore it represents the ideal situation where the EDR space is included in the central subspace.

On several simulation runs, the number of directions chosen by cross-validation MAVE was 2 which is optimal since the functions $\beta_1$ and $\beta_2$ are respectively the second and third eigenvalues of the covariance operator of $X$. Whatever the estimation of the directions $\beta_i$, $i = 1, 2$ are, the quality of prediction is related to how close the estimated projections $\langle \hat{\beta}_i, X \rangle$ are to the true projections $\langle \beta_i, X \rangle$. In order to check this, Figure 6 displays the indexes $\langle \hat{\beta}_i, X \rangle$ versus $\langle \beta_i, X \rangle$ for a typical simulation run, showing a quite satisfactory estimation.

## 5. Conclusion

This survey paper has investigated several aspects of wavelet thresholding and has considered two recent applications of wavelet to solve some interesting statistical problems. We would like however to mention here few more types of signal processing problems where these methods have been used in practice. Wavelet analysis and denoising have been found particularly useful in detecting machinery fault detection. Typical examples of signals encountered in this field are vibration signals generated in defective bearings and gears rotating at constant speeds (see e.g. Lada et al. (2002)). Wavelet denoising procedures in conjunction with hypothesis testing have also been used for detecting change points in several biomedical applications. Typical examples are the detection of life-threatening cardiac arythmias in electrocardiographic signals (ECG) recorded during the monitoring of patiens, or the detection of venous air embolism in doppler heart sound signals recorded during surgery when the incision wounds lie above the heart. The same is true for functional analysis of variance problems and for nonparametric mixed-effects models used in proteomics (e.g. Morris and Carroll (2006), Antoniadis and Sapatinas (2007)). A number of interesting applications of wavelets may be also found in economic and financial applications (e.g. Ramsay and Lampart (1998)) and for times series analysis and prediction (e.g Fryzlewicz et al. (2003), Antoniadis and Sapatinas (2003)). In conclusion, it is apparent that wavelets are particularly well adapted to the statistical analysis of several types of data.

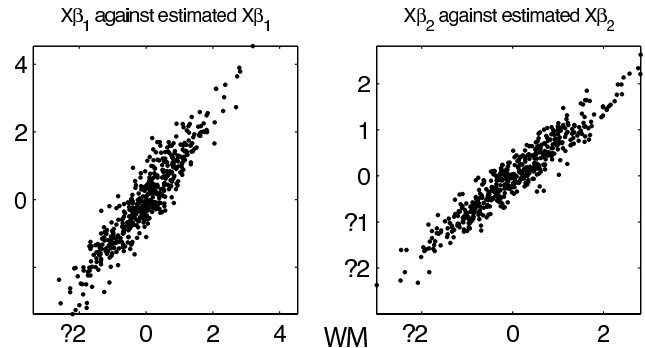

Fig 6. *True projections versus estimated projections for the simulated example with the wavelet MAVE method (WM).*